\documentclass[journal]{IEEEtran}
\usepackage{amssymb,amsmath,theorem}
\usepackage{array}
\usepackage{graphicx}
\usepackage{epsfig}
\usepackage{amsfonts}
\usepackage{bm}
\usepackage{arydshln}
\usepackage{subfig}
\usepackage{algorithm}
\usepackage{algorithmic}
\usepackage{stfloats}
\newtheorem{Theo}{Theorem}

\newcolumntype{C}[1]{>{\centering}p{#1}}
\usepackage{rotating,setspace,latexsym,amsmath,epsf,amssymb,bm,color}
\usepackage{cite}
\usepackage{booktabs}

\usepackage{caption}
\IEEEoverridecommandlockouts

\ifCLASSINFOpdf
\else
\fi

\begin{document}
%

\title{Distributed MIMO Positioning: Fundamental Limit Analysis and User Tracking Framework Design}

%

\author{Yingjie~Xu,~\IEEEmembership{Student member,~IEEE},
Xuesong~Cai,~\IEEEmembership{Senior member,~IEEE}, 
Ali~Al-Ameri,~\IEEEmembership{Student member,~IEEE},
        Sara~Willhammar,~\IEEEmembership{Member,~IEEE}, and~Fredrik~Tufvesson,~\IEEEmembership{Fellow,~IEEE}
\thanks{This work was supported by the Swedish strategic research area ELLIIT, by the Vinnova competence center NextG2Com within the Advanced Digitalisation program, grant number 2023-00541 and also partially funded by 6GTandem, supported by the Smart Networks and Services Joint Undertaking (SNS JU) under the European Union's Horizon Europe research and innovation programme under Grant Agreement No 101096302. Parts of this paper were submitted to IEEE International Conference on Communications~(ICC 2026)~\cite{Y. Xu2025}. \emph{(Corresponding author: Yingjie Xu.)}
	
Y.~Xu, A.~Al-Ameri, S.~Willhammar, and F.~Tufvesson are with the Department of Electrical and Information Technology, Lund University, Lund, Sweden (e-mail: \{yingjie.xu, ali.al-ameri, sara.willhammar, fredrik.tufvesson\}@eit.lth.se).

X. Cai is with the State Key Laboratory of Photonics and
Communications, School of Electronics, Peking University, Beijing, 100871,
P. R. China (email: xuesong.cai@pku.edu.cn).}}

%
%

\markboth{Journal of \LaTeX\ Class Files,~Vol.~xx, No.~x, December~2024}%
{Shell \MakeLowercase{\textit{et al.}}: Bare Demo of IEEEtran.cls for IEEE Journals}
%



\maketitle

\begin{abstract}
    This paper presents a comprehensive study on the 3D  positioning capabilities in distributed multiple-input multiple-output~(MIMO) systems. Unlike previous studies that mainly rely on idealized isotropic antenna models, we adopt a polarimetric model that takes advantage of effective aperture distribution functions to characterize realistic antenna patterns, placements, and polarization effects. Based on this model, we analyze the fundamental limits of UE positioning using the Fisher information matrix (FIM) and the position error bound (PEB). The FIM is shown to be expressed as a weighted sum of the information contributions from individual access point (AP)-UE pairs, with each contribution interpreted geometrically across distance, azimuth, and elevation dimensions. The impact of the UE tilt and the spatial distribution of APs on the PEBs is further analyzed. As a further advancement, we propose a complete positioning framework from a UE tracking perspective. By integrating a global probability hypothesis density filter and a PEB-aware AP management strategy, the framework enables accurate tracking while optimizing AP scheduling. Finally, we present a distributed MIMO channel measurement campaign to validate the proposed framework. The results demonstrate a centimeter-level tracking accuracy. In addition, the PEB-aware AP management strategy is shown to maintain robust tracking performance while significantly reducing the number of concurrently active APs, thus lowering the overall system overhead.
\end{abstract}

%
\IEEEpeerreviewmaketitle

\section{Introduction}

Integrated sensing and communication (ISAC) is envisioned as a key technology for upcoming 6G wireless networks. Beyond traditional information transmission, ISAC extends communication systems to include environmental sensing~\cite{H. Guo2024}. Meanwhile, distributed multiple-input multiple-output~(MIMO), also known as cell-free MIMO~\cite{H. Q. Ngo 2017,E. G. Larsson2014}, is emerging as a new paradigm for 6G MIMO systems. The uniform coverage and spatial diversity enabled by the distributed architecture offer significant improvements in both communication and sensing performance. As a result, integrating distributed MIMO and ISAC into a unified system~\cite{E. C. Strinati2024,U. Demirhan2023 }, 
is a promising direction for further exploring the full potential of ISAC.

From a sensing perspective, an important aspect of ISAC systems is the positioning of user equipment (UE) using radio signals.
For effective system design, it is crucial to understand the maximum achievable positioning accuracy, that is, the fundamental performance limits of one-shot positioning. To this end, a widely used analytical tool is the Cramér-Rao lower bound (CRLB), which provides a tight bound for unbiased position estimates~\cite{Z. Wang2025}. For example, the positioning CRLB based on 5G mmWave signals is derived in~\cite{F. Ghaseminajm2021}.
Additionally, defined as the square root of the CRLB of the UE position, the position error bound (PEB) is introduced in~\cite{Y. Shen2010_1} as a specific metric to quantify positioning limits in general wideband systems. Note that~\cite{F. Ghaseminajm2021,Y. Shen2010_1} focus mainly on co-located MIMO setups. Furthermore, cooperative positioning is addressed in~\cite{Y. Shen2010_2,G. Xia2025}, where PEB is derived for distributed networks. 
More studies on the fundamental limits of one-shot UE positioning can be referred to~\cite{J. Huang2019,C. Park2015,S. Kazemi2020,J. He2024}.    

The aforementioned studies are limited in several aspects. First, most of them focus on analysis using ideal isotropic antenna models. However, non-omnidirectional and non-ideal characteristics of antennas are commonly observed in practice~\cite{M Landmann2004}. Neglecting these characteristics can lead to unrealistic antenna modeling, which further results in inaccurate positioning limit analyzes. Second, most studies (e.g.,~\cite{Z. Wang2025,F. Ghaseminajm2021,Y. Shen2010_1,Y. Shen2010_2}) primarily adopt single-polarized antenna setups, overlooking dual-polarized antennas that are widely deployed in practical systems. Third, the impact of real-world UE behavior on positioning is not considered. For example, among smartphone and wearable/portable internet-of-things users, antenna tilting caused by user posture is common. This has gained increasing attention in system and network design~\cite{3GPP2025}. To address these limitations, the study in~\cite{M Landmann2004} derives the CRLB of angle-of-arrival (AoA) estimation by incorporating the effective aperture distribution function (EADF) of the antenna array. The EADF effectively captures the responses of real-world arrays as it is obtained from measured radiation patterns. 
However, a comprehensive investigation of the positioning accuracy limits under a polarimetric system model, which accounts for realistic array characteristics and practical UE behavior, remains largely unexplored.

In addition to UE one-shot positioning, dynamically tracking mobile UEs is a higher-level application, which can be categorized as a multi-target tracking issue. In this context, random finite set (RFS)-based tracking methods~\cite{W. Wu2023}, which integrate RFS theory and the Bayesian filtering framework, have been widely used. A representative approach is the probability hypothesis density (PHD) filter~\cite{K. Granstrom2012_1,B.-N. Vo2006}, which effectively addresses key challenges during tracking, including uncertain data associations, a time-varying number of targets, missed detections, and false alarms. Based on the PHD filter, various challenges in targeting have been extensively explored in the literature (e.g.~\cite{W. Yi2020,K. Granstrom2012_2}), with a focus on sensor networks. In a mobile communication context, a multimodel PHD filter is designed in~\cite{H. Kim2020_1}, enabling UE tracking using mmWave signals. To further account for multipath effects in communication channels, an extended Kalman PhD filter and a cubature Kalman PHD filter are proposed in~\cite{O. Kaltiokallio2024} and~\cite{H. Kim2020_2}, respectively. However, the approaches in~\cite{H. Kim2020_1, O. Kaltiokallio2024, H. Kim2020_2} are limited to co-located MIMO systems and lack support for distributed MIMO deployments. 

Compared to co-located MIMO, distributed MIMO architectures introduce new challenges for UE tracking. Specifically, due to the limited communication range and computational resources, it is neither efficient nor practical for all APs to continuously communicate with the UEs~\cite{U. Demirhan2023}. Therefore, reliable AP management is essential to determine AP activities at any given time. Field-of-view (FoV)-based methods have been proposed in~\cite{Y. Ge2024, Y. Xu2025}. However, accurately characterizing FoVs becomes challenging when realistic antenna arrays are considered. Error bound strategies have been applied in~\cite{M. Arash2022,R Tharmarasa2007,J.Sun2023}, in which devices with lower tracking error bounds are more likely to be selected. However, most of these strategies are evaluated through simulations, lacking experimental performance validation in real-world environments.

To the best of our knowledge, fully exploring ISAC capabilities of distributed MIMO systems, from the perspectives of both UE (one-shot) positioning and tracking, remains an open problem. In addition, UE positioning and tracking performance evaluations that incorporate realistic antenna characterizations, along with experimental validation, are still lacking in the existing literature. To address these gaps, this paper introduces a polarimetric EADF-based system model and presents a detailed investigation of the performance limits of UE positioning. Furthermore, we propose a PHD-based UE tracking framework, enhanced by a PEB-aware AP management strategy. Finally, we conduct real-world distributed MIMO channel measurements to experimentally evaluate the proposed tracking framework. The main contributions of this paper are summarized as follows.
  \begin{enumerate}
	\item  We present a polarimetric system model for distributed MIMO systems, which incorporates the EADF to characterize realistic array responses and introduces UE tilt.
    The motivation of this model is to make the performance evaluation more representative of the behavior of real-world devices on both the AP and the UE sides.
    The fundamental performance limits of one-shot UE positioning are investigated in terms of the Fisher information matrix (FIM) and the PEB.
	\item Based on the derived FIM and PEB, we analyze the impact of spatial distributions of the APs on positioning. An AP geometry factor is defined to effectively quantify the relationship between AP deployment and positioning performance. In addition, by modeling the random UE tilt angle with the Von Mises distribution, the FIM and PEB are discussed while incorporating the UE tilt. Furthermore, the performance gap between the ideal and our realistic EADF-based antenna models is investigated.
	\item We propose a complete framework for 3D UE tracking. With this framework, a global PHD filter is implemented with a PEB-aware AP management strategy to dynamically determine the APs that are activated. This enables smooth tracking while significantly reducing system overhead. To solve the optimization problem involved in the AP management, an iterative greedy-local-search method is introduced to effectively obtain near-optimal solutions.  
	\item We perform an indoor distributed MIMO channel measurement campaign to validate the proposed tracking framework. The proposed framework achieves a centimeter-level root mean square tracking error. Meanwhile, the PEB-aware AP management strategy significantly reduces the number of concurrently active APs while maintaining robust tracking performance.                    

  \end{enumerate}
  \begin{figure}[tb]
	\centering
		\begin{minipage}[tb]{0.48\textwidth}
			\centering
			\includegraphics[width=1\textwidth]{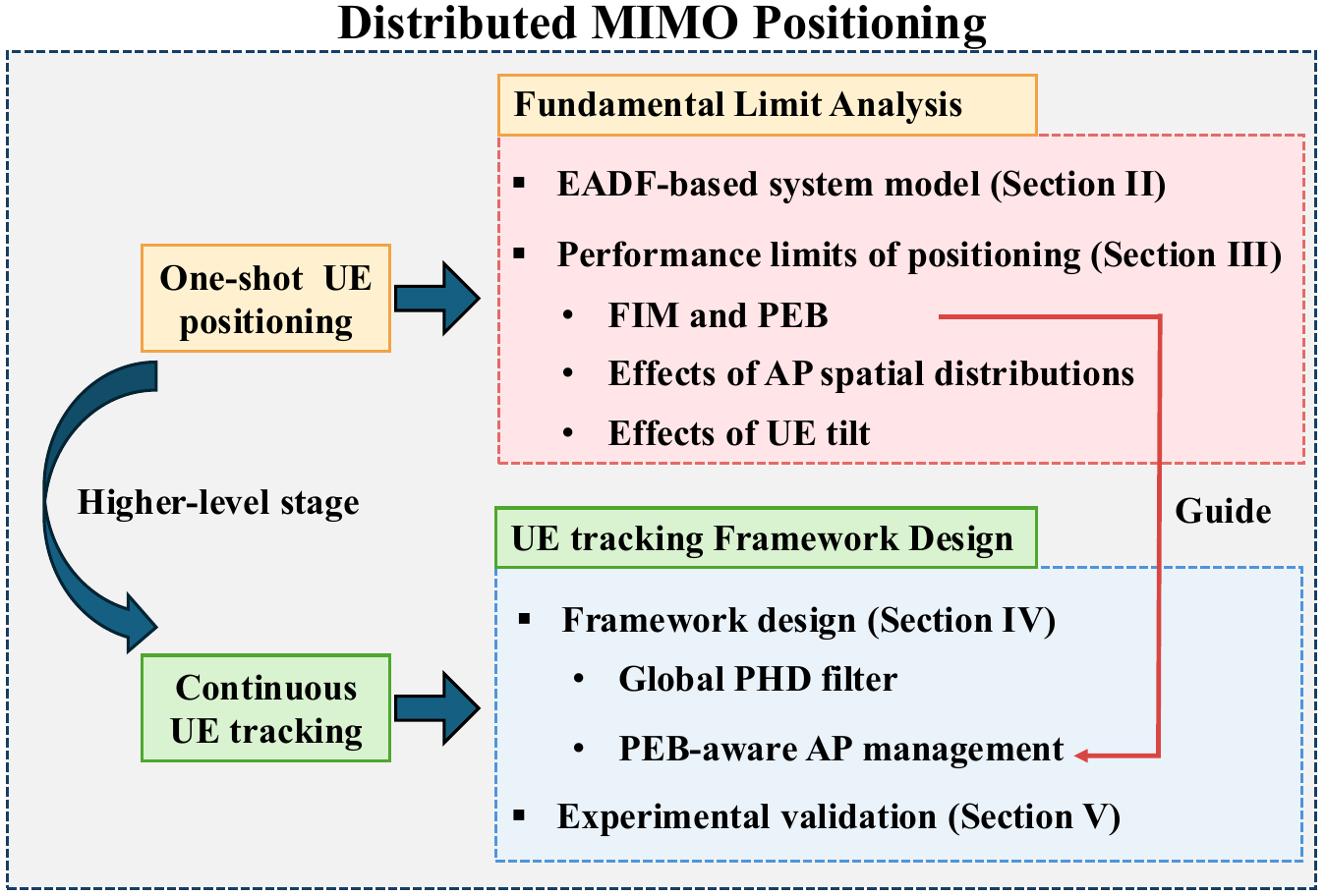}
		\end{minipage}
	\caption{Organization of the paper.}
	\label{fig:paper_organization}
\end{figure}
  The organization of this paper is shown in Fig.~\ref{fig:paper_organization}. Section~II presents the system model. Section~III provides a detailed analysis of the performance limits of UE positioning accuracy in distributed MIMO systems. In Section~IV, the proposed UE tracking framework is introduced. Section~V describes the distributed MIMO channel measurements and the experimental evaluation of the framework. Finally, Section~VI concludes the paper. 

\section{System Model}
Consider a multi-user distributed MIMO system in which $K$ spatially distributed APs serve $M$ users. Each AP is equipped with an antenna array consisting of $N$ antenna elements, while each user is equipped with a single antenna. Antenna tilting may occur on the UE side, we denote the antenna tilt angle of the $m$-th UE in the vertical direction with $\beta_m$. Without loss of generality, we consider uplink communication. The received signals with bandwidth $B$ are sampled with a frequency spacing of $\Delta B$. We assume that all APs and UEs are synchronized. Given the signal $\mathbf{s}_{m}$ transmitted from the $m$-th user, the discrete received signal vector $\mathbf{y}(t_{n_t},f_{n_f})\in \mathbb{C}^{N\times1}$ at the $k$-th AP can be expressed as
\begin{equation}
\begin{split}
    \mathbf{y}(t_{n_t},f_{n_f})=\sum_{l=1}^{L_{k,m}}&\mathbf{C}_{\text{R}}\left(\theta_{k,m}^{(l)},\phi_{k,m}^{(l)}\right)\mathbf{A}_{k,m}^{(l)}\mathbf{r}_{\text{T}}\left[\mathbf{C}_{\text{T},m}\right]^{\text{T}}\mathbf{s}_{m}(f_{n_f})\\
    &\cdot e^{-j2\pi (f_{n_f}\tau_{k,m}^{(l)}+v_{k,m}^{(l)}t_{n_t})}+\mathbf{n}_{k,m},
\end{split}
\end{equation}
where $L_{k,m}$, $\mathbf{n}_{k,m}$, $n_t\in\{1,...,N_t\}$, and $n_f\in\{1,...,N_f=\frac{B}{\Delta B}\}$ are the number of multipath components~(MPCs), the additive Gaussian noise with covariance matrix $\mathbf{R}_{n}=\sigma^2\mathbf{I}_{N\times1}$, and the discrete time and frequency indices, respectively. The terms $\mathbf{C}_{\text{T},m}=\begin{bmatrix} c_{\rm{T_{V}}} & c_{\rm{T_{H}}} \end{bmatrix}\in \mathbb{C}^{1\times2}$ and $\mathbf{C}_{\text{R}}\left(\theta_{k,m}^{(l)},\phi_{k,m}^{(l)}\right)=\begin{bmatrix} \mathbf{c}_{\rm{R_{V}}}\left(\theta_{k,m}^{(l)},\phi_{k,m}^{(l)}\right) & \mathbf{c}_{\rm{R_{H}}}\left(\theta_{k,m}^{(l)},\phi_{k,m}^{(l)}\right) \end{bmatrix}\in \mathbb{C}^{N\times2}$ denote the antenna and array responses at the UE and AP, respectively. Both responses include two orthogonal polarization components rather than assuming idealized polarization separation. Each MPC is parameterized by $\Theta_{l}$, which includes the delay $\tau_{k,m,l}$, azimuth AoA $\phi_{k,m,l}$, elevation AoA $\theta_{k,m,l}$, Doppler frequency $v_{k,m,l}$, and polarization matrix $\mathbf{A}_{k,m,l}$, defined as 
${{\bf{A}}_l} = \left[ {\begin{array}{*{20}{c}}
	{{\alpha _{\rm{VV}}}}&{{\alpha _{\rm{VH}}}}\\
	{{\alpha _{\rm{HV}}}}&{{\alpha _{\rm{HH}}}}
\end{array}} \right] = \left[ {{\alpha _{l,\text{p}\text{p}'}}} \right]$, $\text{p},\text{p}'\in\{V,H\}$. The angles $\phi_{k,m,l}$ and $\theta_{k,m,l}$ are defined in the local coordinate system~(LCS) of each AP. The matrix $\mathbf{r}_{\text{T}}$ characterizes the polarization rotation~\cite{C. Oestges2008} caused by the UE tilt and is given by
$ \mathbf{r}_{\text{T}}=\begin{bmatrix} \cos \beta_m  & \sin\beta_m \\  -\sin\beta_m  & \cos\beta_m \end{bmatrix}.$
By stacking $\mathbf{y}(t_{n_t},f_{n_f})$ from all time and frequency samples, the vectorized received signal $\mathbf{Y}_{k,m}\in \mathbb{C}^{NN_tN_f\times1}$ is obtained as
\begin{equation}
\mathbf{Y}_{k,m}=[\mathbf{y}(t_1,f_1),...,\mathbf{y}(t_{N_t},f_{N_f})]^{\text{T}}= \sum_{l=1}^{L_{k,m}}\mathbf{B}(\mathbf{\Theta}_{l})\cdot \gamma_l +\mathbf{n}_{k,m},
\label{eq:signal model}
\end{equation}
where $\gamma_l = \text{Vec}(\mathbf{A}_l)=\begin{bmatrix} \alpha_{\rm{VV}}  & \alpha_{\rm{VH}}  &  \alpha_{\rm{HV}}  &  \alpha_{\rm{HH}} \end{bmatrix} ^{\text{T}}$, and
\begin{equation}
	\mathbf{B}(\mathbf{\Theta}_l) = \left[
	\begin{array}{c}
	 {c}_{\rm{T_V}}'\left(\mathbf{b}_t \otimes \mathbf{c}_{\rm{R_V}}  \otimes \mathbf{b}_f\right) \\
		 {c}_{\rm{T_H}}'\left(\mathbf{b}_t \otimes \mathbf{c}_{\rm{R_V}} \otimes \mathbf{b}_f\right) \\
		{c}_{\rm{T_V}}' \left(\mathbf{b}_t \otimes \mathbf{c}_{\rm{R_H}}  \otimes \mathbf{b}_f\right) \\
		 {c}_{\rm{T_H}}'\left(\mathbf{b}_t \otimes \mathbf{c}_{\rm{R_H}} \otimes \mathbf{b}_f\right)
	\end{array}
	\right]^{\text{T}} \in \mathbb{C}^{NN_tN_f\times4},
	\label{eq:B}
\end{equation}
with $\mathbf{b}_t=\left[e^{j2\pi v_{k,m}^{(l)}t_1},...,e^{j2\pi v_{k,m}^{(l)}t_{N_t}}\right]^{\text{T}}\in \mathbb{C}^{N_t\times1}$, $\mathbf{b}_f=\left[s_{k,m}(f_1)e^{j2\pi f_1\tau_{k,m}^{(l)}},...,s_{k,m}(f_{N_f})e^{j2\pi f_{N_f}\tau_{k,m}^{(l)}}\right]^{\text{T}}\in \mathbb{C}^{N_f\times1}$, and 
\begin{equation}
\begin{bmatrix} c_{\rm{T_V}}' \\ c_{\rm{T_H}}'  \end{bmatrix}= \begin{bmatrix}  c_{\rm{T_V}}\cos \beta_m - c_{\rm{T_H}}\sin \beta_m\\ c_{\rm{T_V}}\sin \beta_m + c_{\rm{T_H}}\cos \beta_m  \end{bmatrix}.
\label{eq: c'}
\end{equation}

 The EADF has been widely used to obtain full-directional antenna responses based on measured radiation patterns, which are typically sampled in discrete and limited directions~\cite{X. Cai2023}. Specifically, let $\tilde{\mathbf{C}}^{(n)}_{\rm{V/H}}\in\mathbb{C}^{M_{\theta}\times M_{\phi}}$ denote the measured 3D V/H-polarized radiation pattern of the antenna element $n$, where $M_{\theta}$ and $M_{\phi}$ represent the numbers of discrete elevation and azimuth angle samples, respectively. The rows and columns of $\tilde{\mathbf{C}}^{(n)}_{\rm{V/H}}$ correspond to $[0,\frac{2\pi}{M_\theta},...,\frac{2\pi(M_\theta-1)}{M_\theta}]^{\text{T}}$ and $[0,\frac{2\pi}{M_\phi},...,\frac{2\pi(M_\phi-1)}{M_\phi}]^{\text{T}}$, respectively. Defined as the 2D discrete Fourier transform~(2D-DFT) of $\tilde{\mathbf{C}}_{\rm{V/H}}^{(n)}$, the $(m_{\theta},m_{\phi})$-th element of the EADF $\mathbf{Q}_{V/H}^{(n)}\in\mathbb{C}^{M_{\theta}\times M_{\phi}}$ is computed as
$q^{(n)}_{m_{\theta},m_{\phi}}=\bm{\mu}^{\text{T}}_{m_{\theta}}\tilde{\mathbf{C}}^{(n)}_{\rm{V/H}}\bm{\mu}_{m_{\phi}}$,
where
$\bm{\mu}_{m_{\theta}}=\exp\left(-j\pi\frac{2(m_{\theta}-1)-M_{\theta}}{M_{\theta}}[0,...,M_{\theta}-1]^{\text{T}}\right),$
and
$\bm{\mu}_{m_{\phi}}=\exp\left(-j\pi\frac{2(m_{\phi}-1)-M_{\phi}}{M_{\phi}}[0,...,M_{\phi}-1]^{\text{T}}\right).$
Next, by applying the 2D inverse DFT~(IDFT) to $\mathbf{Q}_{V/H}$, the response of the array in an arbitrary direction $(\theta, \phi)$ can be reconstructed as
\begin{equation}
\begin{split}
	\mathbf{c}_{\rm{V/H}}(\theta,\phi) = &\left[c_{\rm{V/H}}^{(1)}(\theta,\phi),...,c_{\rm{V/H}}^{(N)}(\theta,\phi)\right]^{\text{T}} \\
	=& \mathbf{Q}_{\rm{V/H}}\cdot (\mathbf{w}_{\theta}\otimes\mathbf{w}_{\phi}) \\
=&\left[\text{vec}\left(\mathbf{Q}_{\rm{V/H}}^{(1)}\right)^{\text{T}},...,\text{vec}\left(\mathbf{Q}_{\rm{V/H}}^{(N)}\right)^{\text{T}}\right]^{\text{T}}\\
&\cdot (\mathbf{w}_{\theta}\otimes\mathbf{w}_{\phi}),
\label{eq: c_theta_phi}
\end{split}
\end{equation}
where
$\mathbf{w}_{\theta} = \exp\left(j\theta\left[-\frac{M_{\theta}}{2},...,\frac{M_{\theta}}{2}-1\right]\right),$
and
$\mathbf{w}_{\phi} = \exp\left(j\phi\left[-\frac{M_{\phi}}{2},...,\frac{M_{\phi}}{2}-1\right]\right).$

Based on the EADF formulation, the matrix in~(\ref{eq:B}) can be rewritten as
\begin{equation}
	\mathbf{B}(\mathbf{\Theta}_l) = \left[
	\begin{array}{c}
	 {c}_{\rm{T_V}}'\left(\mathbf{b}_t \otimes [\mathbf{Q}_{\rm{R_V}}\cdot(\mathbf{w}_{\theta}\otimes\mathbf{w}_{\phi})]\otimes \mathbf{b}_f\right) \\
		 {c}_{\rm{T_H}}'\left(\mathbf{b}_t \otimes [\mathbf{Q}_{\rm{R_V}}\cdot(\mathbf{w}_{\theta}\otimes\mathbf{w}_{\phi})] \otimes \mathbf{b}_f\right) \\
		{c}_{\rm{T_V}}' \left(\mathbf{b}_t \otimes [\mathbf{Q}_{\rm{R_H}}\cdot(\mathbf{w}_{\theta}\otimes\mathbf{w}_{\phi})]  \otimes \mathbf{b}_f\right) \\
		 {c}_{\rm{T_H}}'\left(\mathbf{b}_t \otimes[\mathbf{Q}_{\rm{R_H}}\cdot(\mathbf{w}_{\theta}\otimes\mathbf{w}_{\phi})] \otimes \mathbf{b}_f\right)
	\end{array}
	\right]^{\text{T}}.
	\label{eq:B2}
\end{equation}
A key advantage of the above EADF-based antenna model is that it is entirely derived from measured radiation patterns~$\tilde{\mathbf{C}}^{(n)}_{\rm{V/H}}$, thus reflecting real-world characteristics of antenna arrays and inherently capturing realistic effects such as manufacturing imperfections, frame asymmetry, and mutual coupling. 

\vspace{-2mm}
\section{Performance Limits of Distributed MIMO Positioning}
This section presents a detailed evaluation of the performance limits of UE positioning accuracy in distributed MIMO systems. Assuming that the locations $\mathbf{p}^{\text{AP}}_k$ and orientations $\mathbf{r}^{\text{AP}}_k$ of the APs are known \emph{a priori}, the objective is to cooperatively localize a UE $m$ with an unknown position $\mathbf{p}^{\text{UE}}_{m}=\left[x_{m}^{\text{UE}},y_{m}^{\text{UE}},z_{m}^{\text{UE}}\right]^{\text{T}}$ based on the received signals $\mathbf{Y}_{k,m}, k=1,...,K$, i.e., one-shot UE positioning. For clarity, we omit the UE index $m$ in this section. We begin by formulating the PEB to characterize the fundamental limits of the positioning accuracy. Then, we analyze the impact of UE tilt and spatial distribution of APs on the PEBs. Finally, numerical simulations are performed to evaluate the achievable positioning accuracy in terms of the PEB.    

\subsection{Position Error Bound}
Let $\hat{\mathbf{p}}^{\text{UE}}$ denote the estimate of ${\mathbf{p}}^{\text{UE}}$. The mean squared error (MSE) matrix of $\hat{\mathbf{p}}^{\text{UE}}$ satisfies the information inequality as 
\begin{equation}
	\mathbb{E} \left\{ ({\mathbf{p}}^{\text{UE}} - \hat{\mathbf{p}}^{\text{UE}})({\mathbf{p}}^{\text{UE}} - \hat{\mathbf{p}}^{\text{UE}})^{\text{T}} \right\} \succeq \mathbf{F}_\mathbf{p}^{-1},
	\label{eq:MSE inequality}
\end{equation}
where $\mathbf{F}_\mathbf{p}$ is the FIM of ${\mathbf{p}}^{\text{UE}}$. Given~(\ref{eq:MSE inequality}), the PEB is then defined as a scalar metric to quantify the lower bound on the positioning RMSE~\cite{Y. Shen2010_1} and is given as
\begin{equation}
\mathcal{P}({\mathbf{p}}^{\text{UE}})\triangleq \sqrt{\text{tr}\{\mathbf{F}_\mathbf{p}^{-1}\}}.
\label{eq:PEB}
\end{equation}
Note that $\mathbf{p}^{\text{UE}}$ is implicitly related to the MPC parameters $\Theta$. We first derive the FIM $\mathbf{F}_{\Theta}$ of $\Theta$, and then formulate the corresponding $\mathcal{P}({\mathbf{p}}^{\text{UE}})$ based on $\mathbf{F}_{\Theta}$. The Doppler frequency is excluded from the estimation parameters as a static scenario is considered. In addition, to obtain an analytical expression of $\mathbf{F}_{\Theta}$, it is a general assumption that MPCs can be well resolved and that the positioning issue is addressed with a single line-of-sight (LoS) path~\cite{Y. Shen2010_1,X. Yin2017}. For the LoS path, assuming that free-space propagation and noise have no effect on signal polarization, the cross-polarization coefficients $\alpha_{VH}$ and $\alpha_{HV}$ in $\mathbf{A}_l$ thus become zero. The signal model in~(\ref{eq:signal model}) can be simplified as
\begin{equation}
\begin{split}
 & \mathbf{Y}_{k,m}= \mathbf{B}(\mathbf{\Theta}_{\text{LoS}})\cdot \gamma_{\text{LoS}} +\mathbf{n}_{k,m}\\
=&\sum_{\text{p}\in\{V,H\}}\alpha_{\text{pp}}\cdot c'_{T_\text{p}}\left(\mathbf{b}_t \otimes [\mathbf{Q}_{R_\text{p}}\cdot(\mathbf{w}_{\theta}\otimes\mathbf{w}_{\phi})] \otimes \mathbf{b}_f\right) +\mathbf{n}_{k,m}.
\label{eq:LoS model}
\end{split}
\end{equation}
In this case, the unknown parameters associated with the $k$-th AP are formulated as (the path index `LoS' is omitted for simplicity)
\begin{equation}
\begin{split}
	\Theta_k = [\underbrace{\theta_k,\phi_k,\tau_k}_{\triangleq \bm{\xi}_k},\underbrace{\mathcal{R}\{\alpha_{\text{VV},k}\},\mathcal{I}\{\alpha_{\text{VV},k}\},\mathcal{R}\{\alpha_{\text{HH},k}\},\mathcal{I}\{\alpha_{\text{HH},k}\}}_{\triangleq \bm{\alpha}_k}].
\end{split}
\end{equation}
The FIM of $\Theta_k$ is expressed as
\begin{equation}
\begin{split}
	\mathbf{F}_{\Theta_k}
	&\triangleq \begin{bmatrix} \mathbf{F}_{\xi\xi} & \mathbf{F}_{\xi\alpha} \\ \mathbf{F}_{\alpha\xi} & \mathbf{F}_{\alpha\alpha} \end{bmatrix} \\
	&=\frac{2}{\sigma^2}\mathcal{R}\left\{ \left(\frac{\partial \mathbf{B}(\mathbf{\Theta}_k)\cdot \gamma_k}{\partial \mathbf{\Theta}_k}\right)^{\text{H}} \cdot \frac{\partial \mathbf{B}(\mathbf{\Theta}_k)\cdot \gamma_k}{\partial \mathbf{\Theta}_k}\right\}.
	\label{eq:F_Theta}
\end{split}
\end{equation}
The diagonal entries for $\theta_k$, $\phi_k$, and $\tau_k$ are expressed as
\begin{equation}
\begin{split}
\left[ \mathbf{F}_{\Theta_k} \right]_{\theta\theta}=\frac{2E_s}{\sigma^2} \mathcal{R}\bigg\{&\left(c_{\rm{T_V}}'\mathbf{A}^{\theta}_{\rm{V}} +c_{\rm{T_H}}'
\mathbf{A}^{\theta}_{\rm{H}}\right)^{\text{H}}\\
&\cdot\left(c_{\rm{T_V}}'\mathbf{A}^{\theta}_{\rm{V}}+c_{\rm{T_H}}'\mathbf{A}^{\theta}_{\rm{H}}\right) \bigg\},
\end{split}
\label{eq:F_thetatheta}
\end{equation}
\begin{equation}
\begin{split}
\left[ \mathbf{F}_{\Theta_k} \right]_{\phi\phi}=\frac{2E_s}{\sigma^2} \mathcal{R}\bigg\{&\left(c_{\rm{T_V}}'\mathbf{A}^{\phi}_{\rm{V}}+c_{\rm{T_H}}'\mathbf{A}^{\phi}_{\rm{H}}\right)^{\text{H}}\\
&\cdot\left(c_{\rm{T_V}}'\mathbf{A}^{\phi}_{\rm{V}}+c_{\rm{T_H}}'\mathbf{A}^{\phi}_{\rm{H}}\right) \bigg\},
\end{split}
\label{eq:F_phiphi}
\end{equation}
\begin{equation}
\begin{split}
\left[ \mathbf{F}_{\Theta_k} \right]_{\tau\tau}=\frac{8\pi^2E_sB_e^2}{\sigma^2} \mathcal{R}\bigg\{&\left(c_{\rm{T_V}}'\mathbf{A}^{\tau}_{\rm{V}}+c_{\rm{T_H}}'\mathbf{A}^{\tau}_{\rm{H}}\right)^{\text{H}}\\
&\cdot\left(c_{\rm{T_V}}'\mathbf{A}^{\tau}_{\rm{V}}+c_{\rm{T_H}}'\mathbf{A}^{\tau}_{\rm{H}}\right) \bigg\},
\end{split}
\label{eq:F_taotao}
\end{equation}
where $E_s=\sum_{n_f=1}^{N_f}\|s(f_{n_f})\|^2$ and $B_e=\frac{\sum_{n_f=1}^{N_f}f_{n_f}\|s(f_{n_f})\|^2}{E_s}$ are the transmission signal power and the effective bandwidth~\cite{Y. Shen2010_1}, respectively. The notations $\mathbf{A}_{\rm{V/H}}^{\theta}$, $\mathbf{A}_{\rm{V/H}}^{\phi}$, and $\mathbf{A}_{\rm{V/H}}^{\tau}$ are expressed as
\begin{equation}
\mathbf{A}_{\rm{V/H}}^{\theta} = \alpha_{\rm{VV/HH}}\cdot \left(\mathbf{Q}_{\rm{R_{V/H}}}\cdot \left[(\mathbf{\alpha}_{\theta}\odot\mathbf{w}_{\theta})\otimes\mathbf{w}_{\phi}\right]\right),
\label{eq:A_theta_V/H}
\end{equation}
\begin{equation}
\mathbf{A}_{\rm{V/H}}^{\phi} = \alpha_{\rm{VV/HH}}\cdot \left(\mathbf{Q}_{\rm{R_{V/H}}}\cdot \left[\mathbf{w}_{\theta}\otimes(\mathbf{\alpha}_{\phi}\odot\mathbf{w}_{\phi})\right]\right),
\label{eq:A_phi_V/H}
\end{equation}
\begin{equation}
\mathbf{A}_{\rm{V/H}}^{\tau} = \alpha_{\rm{VV/HH}}\cdot \left(\mathbf{Q}_{\rm{R_{V/H}}}\cdot \left[\mathbf{w}_{\theta}\otimes\mathbf{w}_{\phi}\right]\right).
\label{eq:A_tau_V/H}
\end{equation}
For a more detailed derivation and other entries, refer to Supplementary Material~S.I in~\cite{Y. Xu2025_2}.

As only the parameters $\theta_k$, $\phi_k$, and $\tau_k$ are of interest to determine $\mathbf{p}^{\text{UE}}$, we adopt the notion of equivalent FIM (EFIM) of $\mathbf{F}_{\xi_k}$~\cite{Y. Shen2010_2}, defined as $\mathbf{F}_{\bm{\xi}_k}=\mathbf{F}_{\xi\xi}-\mathbf{F}_{\xi\alpha}\mathbf{F}_{\alpha\alpha}\mathbf{F}_{\xi\alpha}^{\text{T}}$. The FIM of $\mathbf{p}^{\text{UE}}$ from the $k$-th AP, under its LCS, can then be obtained using the chain rule~\cite{Y. Shen2007} as
\begin{equation}
\mathbf{F}_{\mathbf{p},k}^{\text{Local}}=\mathbf{J}\mathbf{F}_{\bm{\xi}_k}\mathbf{J}^{\text{T}},
\label{eq:F_p_local}
\end{equation}
with the Jacobian matrix $\mathbf{J}=\frac{\partial \bm{\xi}_k^{\text{T}}}{\partial \mathbf{p}^{\text{UE}}}$. Assuming that the measurements from different APs are statistically independent, the joint FIM of the UE position in the global coordinate system~(GCS) is given by
\begin{equation}
\mathbf{F}_{\mathbf{p}}=\sum_{k=1}^{K}\mathbf{r}_k^{\text{AP}}\mathbf{F}_{\mathbf{p},k}^{\text{Local}}(\mathbf{r}_k^{\text{AP}})^{\text{T}},
\label{eq:F_p_joint}
\end{equation}
where we recall that $\mathbf{r}_k^{\text{AP}}$ is the orientation of the AP~$k$.
By substituting the expressions of $\mathbf{F}_{\Theta_k}$ and $\mathbf{F}_{\bm{\xi}_k}$ into~(\ref{eq:F_p_local}) and~(\ref{eq:F_p_joint}), we derive the following theorems.

\begin{Theo}
	Given the FIM entries $\left[ \mathbf{F}_{\Theta_k} \right]_{\theta\theta}$, $\left[ \mathbf{F}_{\Theta_k} \right]_{\phi\phi}$, and $\left[ \mathbf{F}_{\Theta_k} \right]_{\tau\tau}$, the resulting FIM of $\mathbf{p}^{\mathrm{UE}}$, with contributions from the $k$-th AP and defined in its LCS, can be derived as
\begin{equation}
\begin{split}
   \mathbf{F}_{\mathbf{p},k}^{\mathrm{Local}}\approx& \frac{\left[ \mathbf{F}_{\Theta_k} \right]_{\theta\theta}}{c^2\tau_k^2}\mathbf{U}\left(\theta_k+\frac{\pi}{2},\phi_k\right)+\frac{\left[ \mathbf{F}_{\Theta_k} \right]_{\tau\tau}}{c^2}\mathbf{U}\left(\theta_k,\phi_k\right)\\
   &+\frac{\left[ \mathbf{F}_{\Theta_k} \right]_{\phi\phi}}{c^2\tau_k^2\sin^2\theta_k}\mathbf{U}\left(\theta_k,\phi_k+\frac{\pi}{2}\right),
   \label{eq:F_local}
\end{split}
\end{equation}
where $\mathbf{U}(\theta,\phi)$ is the 3D ranging direction matrix (RDM)~\cite{Y. Han2016} defined as
\begin{equation}
\mathbf{U}(\theta,\phi) \triangleq \begin{bmatrix} \cos \phi \sin \theta  \\ \sin \phi \sin \theta  \\ \cos \theta  \end{bmatrix}  \cdot\begin{bmatrix} \cos \phi \sin \theta  \\ \sin \phi \sin \theta  \\ \cos \theta  \end{bmatrix}^{\mathrm{T}}.
\end{equation}
\end{Theo}

\emph{Proof}: Refer to Supplementary Material~S.II in~\cite{Y. Xu2025_2}.

\begin{figure}[tb]
	\centering
		\begin{minipage}[tb]{0.45\textwidth}
			\centering
			\includegraphics[width=1\textwidth]{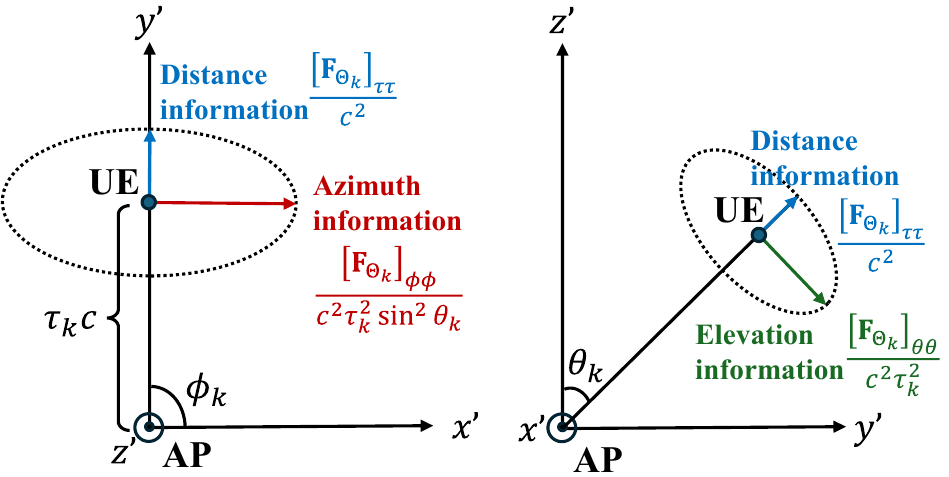}
		\end{minipage}
	\caption{Geometric interpretation of the FIM $\mathbf{F}_{\mathbf{p},k}^{\mathrm{Local}}$: Information ellipse formed by distance, azimuth, and elevation information.}
	\label{fig:information_ellipse}
\end{figure}
From~(\ref{eq:F_local}), the total FIM $\mathbf{F}_{\mathbf{p},k}^{\text{Local}}$ is expressed as a weighted matrix sum of three components, each aligned with an orthogonal direction in a 3D space. First, the term $\frac{\left[\mathbf{F}_{\Theta_k} \right]_{\tau\tau}}{c^2}\mathbf{U}(\theta,\phi)$ reflects the contribution of information from the signal propagation delay/distance, that is, the \emph{distance information}. It is aligned with the direct path between the AP and the UE, as shown in Fig.~\ref{fig:information_ellipse}. This term dominates the FIM when the effective bandwidth is large, since $\frac{\left[\mathbf{F}_{\Theta_k} \right]_{\tau\tau}}{c^2}$ is proportional to $B_e$.
Furthermore, the component $\frac{\left[ \mathbf{F}_{\Theta_k} \right]_{\theta\theta}}{c^2\tau_k^2}\mathbf{U}\left(\theta_k+\frac{\pi}{2},\phi_k\right)$ corresponds to the \emph{elevation information}, while $\frac{\left[ \mathbf{F}_{\Theta_k} \right]_{\phi\phi}}{c^2\tau_k^2\sin^2\theta_k}\mathbf{U}\left(\theta_k,\phi_k+\frac{\pi}{2}\right)$ represents the \emph{azimuth information}. Both are orthogonal to the propagation direction and indicate the system's ability to resolve the angular position of the UE. These components of angular information degrade with increasing propagation distance $d_k=c\tau_k$, leading to a reduced maximum achievable positioning accuracy in both azimuth and elevation directions for more distant UEs with fixed $\left[\mathbf{F}_{\Theta_k} \right]_{\tau\tau}$ and $\left[ \mathbf{F}_{\Theta_k} \right]_{\theta\theta}$. Moreover, the azimuth information is scaled by $\sin^{-2}\theta_k$, which means that the maximum azimuth information $\frac{\left[ \mathbf{F}_{\Theta_k} \right]_{\phi\phi}}{c^2\tau_k^2}$ occurs when the UE lies in the horizontal plane, if other parameters are fixed. 

\begin{Theo}
Assume that the orientation $\mathbf{r}_k^{\text{AP}}$ of the AP $k$ causes the x-axis of its LCS to deviate by an angle $\omega_k$ counterclockwise with respect to the x-axis of the GCS. The joint FIM of $\mathbf{P}^{\text{UE}}$ is derived as
\begin{equation}
\begin{split}
   \mathbf{F}_{\mathbf{p}}\approx&\sum_{k=1}^{K} \frac{\left[ \mathbf{F}_{\Theta_k} \right]_{\theta\theta}}{c^2\tau_k^2}\mathbf{U}\left(\theta_k+\frac{\pi}{2},\phi_k+\omega_k\right)\\
   &+\frac{\left[ \mathbf{F}_{\Theta_k} \right]_{\phi\phi}}{c^2\tau_k^2\sin^2\theta_k}\mathbf{U}\left(\theta_k,\phi_k+\omega_k+\frac{\pi}{2}\right)\\
   &+\frac{\left[ \mathbf{F}_{\Theta_k} \right]_{\tau\tau}}{c^2}\mathbf{U}\left(\theta_k,\phi_k+\omega_k\right).
\end{split}
\label{eq:F_global}
\end{equation}
\end{Theo}
\emph{Proof}: Based on Theorem~1, (\ref{eq:F_global}) can be easily obtained through $\mathbf{F}_{\mathbf{p}} = \sum_{k=1}^{K} \mathbf{F}^{\text{Global}}_{\mathbf{p},k}=\sum_{k=1}^{K} \mathbf{r}^{\text{AP}}_k \mathbf{F}_{\mathbf{p},k}^{\text{Local}} (\mathbf{r}^{\text{AP}}_k)^{\text{T}}$ with
$	\mathbf{r}^{\text{AP}}_k = \begin{bmatrix}
	\cos \omega_k & \sin \omega_k & 0 \\
	-\sin \omega_k & \cos \omega_k & 0 \\
	0 & 0 & 1
	\end{bmatrix}.$

Eq.~(\ref{eq:F_global}) indicates that the joint FIM of $\mathbf{P}^{\text{UE}}$ is a weighted sum of the information contributed by each AP–UE pair. Each pair provides information along the propagation direction, azimuth, and elevation dimensions, which together constitute the complete positioning FIM. This observation agrees with the result of~\cite{Y. Han2016}, although the study therein assumes 2D scenarios and idealized isotropic antenna models.

\setcounter{equation}{25}  
\begin{figure*}[!b]
\hrulefill
\begin{equation}
\mathcal{P}(\mathbf{p}^{\text{UE}})=\sqrt{\frac{4c^2\sum_{k=1}^{K}\left(\frac{[\mathbf{F}_{\Theta_k}]_{\phi\phi} }{\tau_k^2} +[\mathbf{F}_{\Theta_k}]_{\tau\tau}\right)}{ \left|\sum_{k=1}^{K}   \frac{[\mathbf{F}_{\Theta_k}]_{\phi\phi} }{\tau_k^2} -[\mathbf{F}_{\Theta_k}]_{\tau\tau}     \right|^2+   \sum_{k=1}^{K}\sum_{k'=1}^{K} \bigg(\frac{[\mathbf{F}_{\Theta_k}]_{\phi\phi} }{\tau_k^2} [\mathbf{F}_{\Theta_{k'}}]_{\tau\tau} + \frac{[\mathbf{F}_{\Theta_{k'}}]_{\phi\phi} }{\tau_{k'}^2} [\mathbf{F}_{\Theta_k}]_{\tau\tau}\bigg)   }}.
\label{eq:PEB3}
\end{equation}
\end{figure*}
\subsection{Impact of Spatial Distribution of APs }
In this subsection, the impact of the spatial distribution of APs on the UE position error bounds is analyzed. To simplify the problem, we assume a scenario with azimuth-only signal propagation, where the 2D positions of the APs are given by $\mathbf{p}^{\text{AP}}_k=[x^{\text{AP}}_k, y^{\text{AP}}_k]$. Consequently, elevation information $\frac{\left[ \mathbf{F}_{\Theta_k} \right]_{\theta\theta}}{c^2\tau_k^2}\mathbf{U}\left(\theta_k+\frac{\pi}{2},\phi_k\right)$ in~(\ref{eq:F_local}) is omitted, and (\ref{eq:F_global}) is simplified as
\setcounter{equation}{23}  
\begin{equation}
\begin{split}
	\mathbf{F}_{\mathbf{p}}\approx&\sum_{k=1}^{K}\frac{\left[ \mathbf{F}_{\Theta_k} \right]_{\phi\phi}}{c^2\tau_k^2}\tilde{\mathbf{U}}\left(\phi_k+\omega_k+\frac{\pi}{2}\right)+\frac{\left[ \mathbf{F}_{\Theta_k} \right]_{\tau\tau}}{c^2}\tilde{\mathbf{U}}\left(\phi_k+\omega_k\right),
\end{split}
\label{eq:2D F_global}
\end{equation}
where $\tilde{\mathbf{U}}(\phi)\triangleq[\cos\phi\quad\sin\phi]^{\text{T}}\cdot[\cos\phi\quad\sin\phi]$.

Our target is to minimize the PEB through optimizing AP distributions~$[\mathbf{p}^{\text{AP}}_1,...,\mathbf{p}^{\text{AP}}_K]$. Clearly, $\mathbf{F}_{\mathbf{p}}$ is affected by both the propagation distance (which is determined by the AP distributions) and the AP orientations. To derive an analytical expression for the optimal AP distribution, we keep the propagation distance and the angle between the UE and the orientation of each AP fixed. This implies that once an AP's position is determined, $w_k$ (which is not part of the optimization objective) is also fixed. Under these assumptions, we derive the following theorem.

\begin{Theo}
	 Given fixed propagation distances and fixed orientations of the APs, the optimal AP distribution satisfies 
	 \begin{equation}
\sum_{k=1}^{K}\left(  \frac{[\mathbf{F}_{\Theta_k}]_{\phi\phi} }{\tau^2c^2}- \frac{[\mathbf{F}_{\Theta_k}]_{\tau\tau} }{c^2} \right)\cdot e^{j2(\phi_k+\omega_k)}=0.
\label{eq:optimal AP distribution}
\end{equation}
The resulting optimal PEB has a closed-form expression, as shown in~(\ref{eq:PEB3}) at the bottom of this page.

\end{Theo}

\emph{Proof}: Refer to Supplementary Material~S.III in~\cite{Y. Xu2025_2}.

From~(\ref{eq:optimal AP distribution}), we observe that an optimal AP distribution is achieved when the contributions $\left(  \frac{[\mathbf{F}_{\Theta_k}]_{\phi\phi} }{\tau^2c^2}- \frac{[\mathbf{F}_{\Theta_k}]_{\tau\tau} }{c^2} \right)$ of all APs, each weighted by the direction $(\phi_k+\omega_k)$, cancel each other out. Intuitively, a diverse geometric distribution of APs is more likely to satisfy this condition and lead to a smaller PEB. A special case occurs when $  \frac{[\mathbf{F}_{\Theta_k}]_{\phi\phi} }{\tau^2c^2}= \frac{[\mathbf{F}_{\Theta_k}]_{\tau\tau} }{c^2}$, in which the AP distribution has no impact on positioning performance. This is because the distance and azimuth information from each AP is identical, resulting in isotropic positioning~\cite{Y. Han2016}. At this time, the associated information ellipse shown in Fig.~\ref{fig:information_ellipse} becomes a circle, which means that the position of the UE is equally well estimated in all directions, regardless of the AP distribution.

Based on Theorem~3 and inspired by~\cite{Y. Han2016}, we define an \emph{AP geometry factor} as
\setcounter{equation}{26}  
\begin{equation}
D\left(\mathbf{p}^{\text{AP}}_1,...,\mathbf{p}^{\text{AP}}_K\right)=\sum_{k=1}^{K}\left(  \frac{[\mathbf{F}_{\Theta_k}]_{\phi\phi} }{\tau^2c^2}- \frac{[\mathbf{F}_{\Theta_k}]_{\tau\tau} }{c^2} \right)\cdot e^{j2(\phi_k+\omega_k)}
\end{equation}
to quantitatively assess the influence of the AP deployment on the PEB. This metric will be evaluated numerically in Section~III-D.

\subsection{Impact of UE Tilt }
In this subsection, we evaluate the UE positioning bounds by incorporating the impact of the UE tilt, which is quite common in real scenarios and especially among smartphone users. As defined in Section~II, the UE tilt is modeled by a random antenna tilt angle $\beta_m$. To derive an analytical PEB expression in terms of $\beta_m$, we consider a simplified scenario where the UE is equipped with a single-polarized omnidirectional antenna, i.e., $c_{\rm{R_V}}=1$ while $c_{\rm{R_H}}=0$. Then~(\ref{eq: c'}) can be simplified to $\begin{bmatrix} c_{\rm{R_V}}'  & c_{\rm{R_H}}'  \end{bmatrix}^{\text{T}}= \begin{bmatrix}  \cos \beta_m  &\sin \beta_m  \end{bmatrix}^{\text{T}}$. Several distributions have been widely used to model random angles, e.g., normal, Laplacian, and Von Mises distributions. Among them, the Von Mises distribution is adopted here due to its natural suitability for circular random variables~\cite{W. Queiroz2011}. Specifically, we assume that $\beta_m$ follows a Von Mises (VM) distribution with
\begin{equation}
\beta_m \sim \mathcal{VM}(\mu_m,\kappa_m)=\frac{e^{\kappa_m \cos(\beta_m-\mu_m)}}{2\pi I_0(\kappa_m)}, \quad \beta_m \in [0, 2\pi],
\label{eq:beta_vonmises}
\end{equation}
where $\mu_m$ and $\kappa_m$ are the mean and concentration parameters, and $I_0(\cdot)$ denotes the modified Bessel function of the first kind. We further define an averaged FIM of ${\Theta_k}$ as $\bar{\mathbf{F}}_{\Theta}=\mathbb{E}\{{\mathbf{F}}_{\Theta}\}=\left[ \mathbb{E}\{ [\mathbf{F}_{\Theta}]_{zz'} \} \right]$, $z,z'\in\Theta_k$. Using this definition, we establish the following theorem.

\begin{Theo}
	 Assume $\beta_m$ follows the Von Mises distribution in~(\ref{eq:beta_vonmises}) and the averaged FIM $\bar{\mathbf{F}}_{\mathbf{p}}$ is obtained by replacing $\left[ \mathbf{F}_{\Theta_k} \right]_{zz'}$ in~(\ref{eq:F_global}) with its expectation $\mathbb{E}\left\{ \left[ \mathbf{F}_{\Theta_k} \right]_{zz'}\right\}$. Furthermore, under the assumption that $\left(\mathbf{c}_{\rm{V}}\right) ^{\mathrm{H}}\mathbf{c}_{\rm{H}}+\left(\mathbf{c}_{\rm{H}}\right) ^{\mathrm{H}}\mathbf{c}_{\rm{V}}\ll \left(\mathbf{c}_{\rm{V}}\right) ^{\mathrm{H}}\mathbf{c}_{\rm{V}}+\left(\mathbf{c}_{\rm{H}}\right) ^{\mathrm{H}}\mathbf{c}_{\rm{H}}$, where $\mathbf{c}_{\rm{V/H}}$ is defined in~(\ref{eq: c_theta_phi}), $\mathbb{E}\left\{ \left[ \mathbf{F}_{\Theta_k} \right]_{zz'}\right\}$ has the following analytical expression:
	 \begin{equation}
     \begin{split}
\mathbb{E}\left\{ \left[ \mathbf{F}_{\Theta_k} \right]_{zz'}\right\}=\frac{2E_s}{\sigma^2} \mathcal{R}\bigg\{&\frac{1+\rho}{2}(\mathbf{A}_{\rm{V}}^{z})^{\mathrm{H}}\mathbf{A}_{\rm{V}}^{z'}\\
&+\frac{1-\rho}{2}(\mathbf{A}_{\rm{H}}^{z})^{\mathrm{H}}\mathbf{A}_{\rm{H}}^{z'}\bigg\},
\end{split}
\label{eq:F_zz}
\end{equation}
where $\rho=\frac{I_2(\kappa_m)}{I_0(\kappa_m)}\cos(2\mu_m)$. The terms $\mathbf{A}_{\rm{V/H}}^{\theta}$, $\mathbf{A}_{\rm{V/H}}^{\phi}$, and $\mathbf{A}_{\rm{V/H}}^{\tau}$ have been defined in~(\ref{eq:A_theta_V/H})--(\ref{eq:A_tau_V/H}), while for the other terms, 
\begin{equation}
\mathbf{A}_{\mathrm{p}}^{\mathcal{R}\{\alpha_{\mathrm{pp}}\}}=\alpha_{\mathrm{pp}}\left(\mathbf{Q}_{\rm{R}_{\mathrm{p}}}\cdot \left[\mathbf{w}_{\theta}\otimes\mathbf{w}_{\phi}\right]\right),
\end{equation}
\begin{equation}
\mathbf{A}_{\mathrm{p}}^{\mathcal{I}\{\alpha_{\mathrm{pp}}\}}=j\alpha_{\mathrm{pp}}\left(\mathbf{Q}_{\rm{R}_{\mathrm{p}}}\cdot \left[\mathbf{w}_{\theta}\otimes\mathbf{w}_{\phi}\right]\right),
\end{equation}
where $\mathrm{p}\in \{V,H\}$, and $\mathbf{A}_{H}^{\mathcal{R}\{\alpha_{\mathrm{VV}}\}}=\mathbf{A}_{H}^{\mathcal{I}\{\alpha_{\mathrm{VV}}\}}=\mathbf{A}_{V}^{\mathcal{R}\{\alpha_{\mathrm{HH}}\}}=\mathbf{A}_{V}^{\mathcal{I}\{\alpha_{\mathrm{HH}}\}}=0$. 
\end{Theo}

\emph{Proof}: Refer to Supplementary Material~S.IV in~\cite{Y. Xu2025_2}.

In Theorem~4, the assumed inequality between $c_{V}$ and $c_{H}$ reflects the scenario in which the cross-polarized antenna responses are much weaker than the co-polarized responses, representing an antenna with high cross-polarization isolation performance~(XPI)~\cite{C. Oestges2008}. Based on Theorem~4, we further define an averaged EFIM $\bar{\mathbf{F}}_{\xi_k}$ as $\bar{\mathbf{F}}_{\xi_k}=\bar{\mathbf{F}}_{\xi\xi}-\bar{\mathbf{F}}_{\xi\alpha}\bar{\mathbf{F}}_{\alpha\alpha}\bar{\mathbf{F}}_{\xi\alpha}^{\text{T}}$, where $\bar{\mathbf{F}}_{\xi\xi}$, $\bar{\mathbf{F}}_{\xi\alpha}$, and $\bar{\mathbf{F}}_{\alpha\alpha}$ are the sub blocks of $\bar{\mathbf{F}}$ (see~(\ref{eq:F_Theta})). The averaged PEB is then given by 
\begin{equation}
\bar{\mathcal{P}}({\mathbf{p}}^{\text{UE}})\triangleq \sqrt{\text{tr}\{\bar{\mathbf{F}}_\mathbf{p}^{-1}\}}=\sqrt{\text{tr}\left\{\left[\sum_{k=1}^{K} \mathbf{J}\bar{\mathbf{F}}_{\xi_k}\mathbf{J}^{\text{T}}\right]^{-1} \right\}},
\end{equation}
which will be evaluated numerically in the next subsection.

\subsection{Numerical Results}
\begin{figure}[t]
	\centering
	\subfloat[]
	{
		\begin{minipage}[tb]{0.25\textwidth}
			\centering
			\includegraphics[width=1\textwidth]{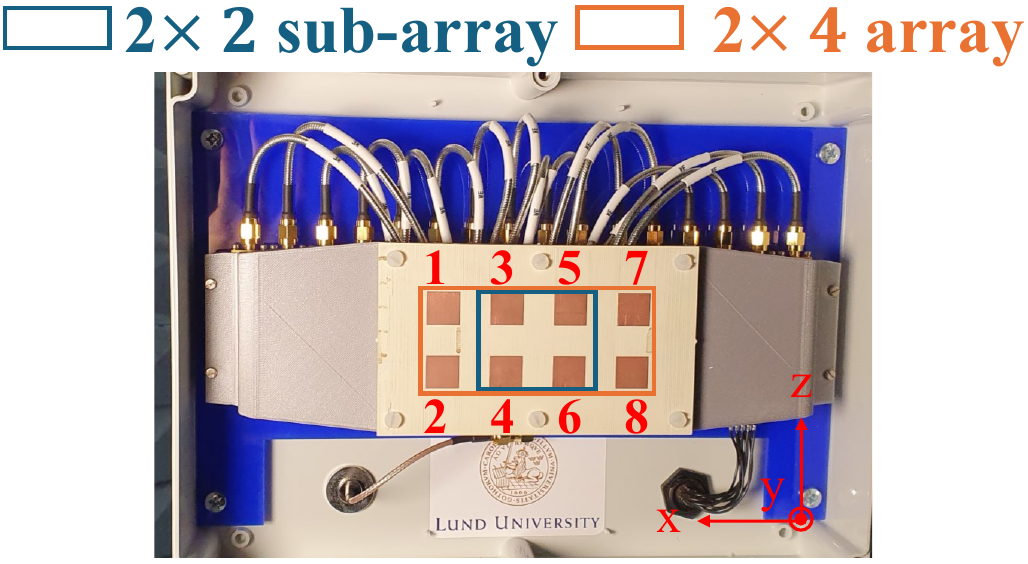}
		\end{minipage}
	}
		\subfloat[]
	{
		\begin{minipage}[tb]{0.25\textwidth}
			\centering
			\includegraphics[width=1\textwidth]{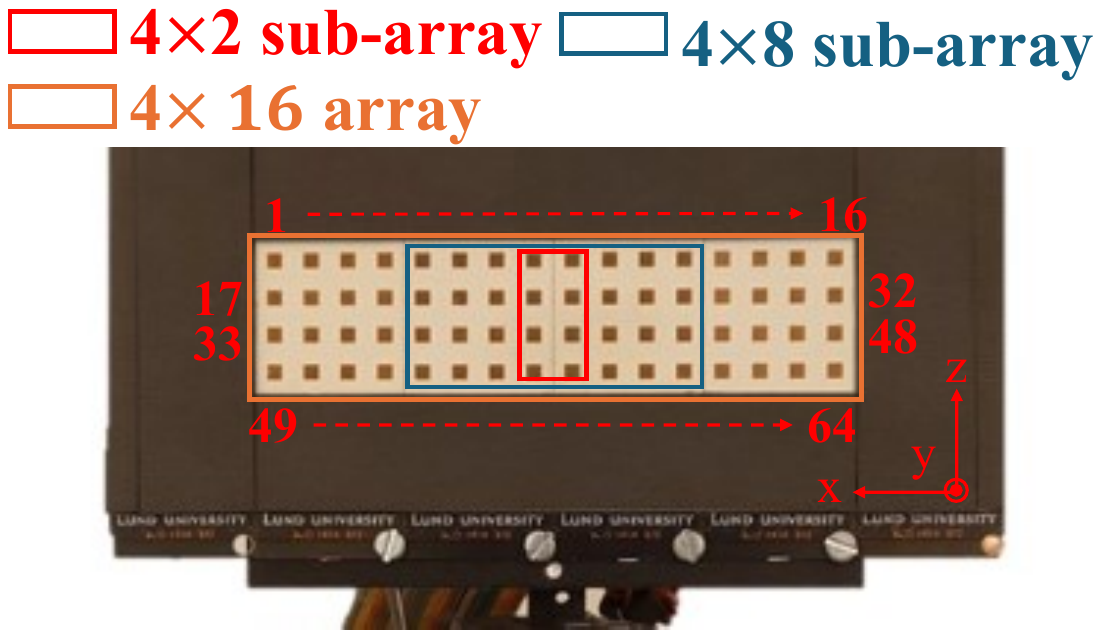}
		\end{minipage}
	}
	
	\caption{Photos of (a) the sub-6~GHz UPA and (b) the mmWave UPA. Note that each patch element has two ports with different polarizations.}
	\label{fig:UPA photo}
\end{figure}

In this subsection, numerical results are provided to evaluate the UE positioning performance. The AP-side antenna response is characterized by the measured EADF of two real-world arrays. The first array is a $2\times4$ dual-polarized uniform planar array (UPA) designed for the 5.0--6.0~GHz frequency band (sub-6~GHz), as shown in Fig.~\ref{fig:UPA photo}a. The second is a $4\times16$ dual-polarized UPA designed for the 26--30~GHz (mmWave) band, as shown in Fig.~\ref{fig:UPA photo}b. For detailed antenna properties, please refer to~\cite{M. Sandra2024, X. Cai2024}. The parameter settings used for the EADF measurements are summarized in Table~\ref{tab:UPA parameters}. Note that the EADF is normalized after the measurements. Specifically, the measured EADF $\mathbf{Q}_{\text{p},m_\theta,m_\phi}^{(n)}$, $\text{p}\in\{V,H\}, m_\theta\in [1,M_\theta], m_\phi\in [1,M_\phi]$, for the $n$-th element with the $n_f$-th discrete frequency, is normalized as
\begin{equation}
\tilde{\mathbf{Q}}_{\text{p},m_\theta,m_\phi}^{(n,n_f)}=\frac{\mathbf{Q}_{\text{p},m_\theta,m_\phi}^{(n,n_f)}}{\sqrt{\sum_{n_f=1}^{N_f} \sum_{m_\theta=1}^{M_\theta}\sum_{m_\phi=1}^{M_\phi}\sum_{\text{p}\in\{V,H\}}\left\|\mathbf{Q}_{\text{p},m_\theta,m_\phi}^{(n,n_f)}\right\|^2}}.
\end{equation}
This operation normalizes the expected energy of each element to one while preserving the intra-element variations across frequencies, polarizations, and directions, as well as the inter-element amplitude differences. Unless otherwise specified, we assume a single vertically polarized UE in the simulations with antenna responses $c_{\rm{R_V}}=1$ and $c_{\rm{R_H}}=0$, and a tilt angle of $\beta=45^{\circ}$. The channel amplitude $
\alpha_{\rm{VV}}$ is modeled based on the free-space path loss, given by $\alpha_{\rm{VV}}=\frac{\lambda}{4\pi \|\mathbf{P}^{\text{AP}}-\mathbf{P}^{\text{UE}}\|}$. The transmission signal power is set to $E_s=\frac{|\alpha_{\rm{VV}}|^2}{NN_f}$. We define the receive signal-to-noise ratio (SNR) as $\frac{E_s}{\sigma^2}$.

\begin{table}[tb]
\centering
\footnotesize
\caption{Parameters of the measured EADF}
\label{tab:UPA parameters}
\begin{tabular}{>{\centering\arraybackslash}m{3.8cm} >{\centering\arraybackslash}m{1.9cm} >{\centering\arraybackslash}m{1.9cm}}
\toprule
 & Sub-6~GHz UPA & mmWave UPA \\
\midrule
Number of antenna elements $N$ & 8 (16 ports) & 64 (128 ports) \\
Measured frequency band (GHz) & 5.4--5.8 & 27.6--28.4 \\

Frequency samples $N_f$ & 32 & 41 \\
Elevation angle samples $M_{\theta}$ & 36 & 60 \\
Azimuth angle samples $M_{\phi}$ & 36 & 60 \\
\bottomrule
\end{tabular}
\end{table}

\begin{figure}[t]
	\centering
	\subfloat[]
	{
		\begin{minipage}[tb]{0.24\textwidth}
			\centering
			\includegraphics[width=1\textwidth]{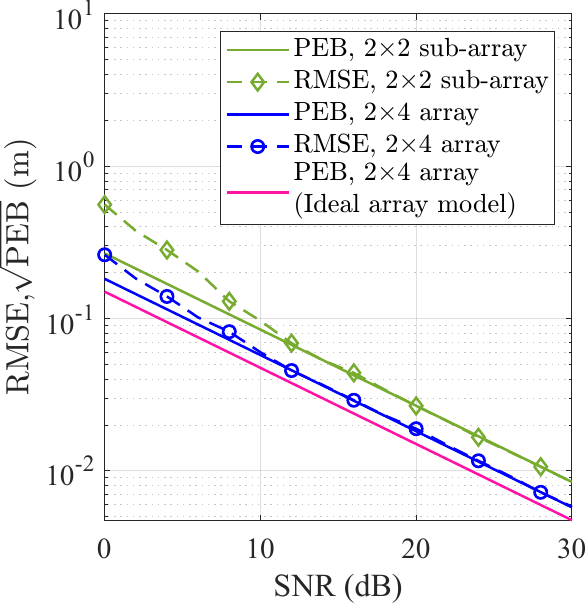}
		\end{minipage}
	}
		\subfloat[]
	{
		\begin{minipage}[tb]{0.24\textwidth}
			\centering
			\includegraphics[width=1\textwidth]{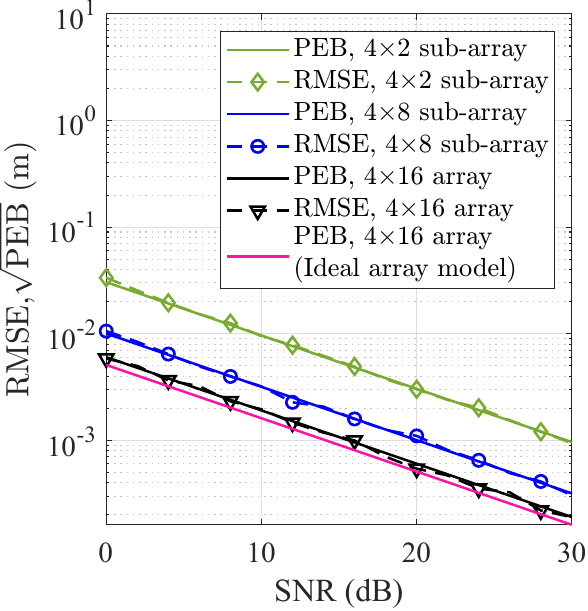}
		\end{minipage}
	}
	
	\caption{Comparisons between the derived PEBs and RMSEs achieved by a maximum likelihood estimator with (a) the sub-6~GHz UPA and (b) the mmWave UPA.}
	\label{fig:PEBvsArraySize}
\end{figure}

\subsubsection{Impact of Array Aperture}We consider a single AP located at $\mathbf{p}^{\text{AP}}=[0,0,0]^{\text{T}}$ with orientation $\mathbf{r}^{\text{AP}}= \mathbf{I}_{3\times3}$. The UE is fixed at $\mathbf{p}^{\text{UE}}=[2,2,2]^{\text{T}}.$ Here, the root mean squared error (RMSE) of the estimated UE position, achieved by a maximum likelihood estimator, is also evaluated compared to the derived PEB. The RMSE is defined as 
$\text{RMSE}(\mathbf{p}^{\text{UE}})=\mathbb{E}\left\{\sqrt{\|\hat{\mathbf{p}}^{\text{UE}}-\mathbf{p}^{\text{UE}}\|^2}\right\}$,
where $\hat{\mathbf{p}}^{\text{UE}}$ is the estimated UE position. The numerical RMSE is obtained from 1000 Monte Carlo experiments.

By selecting different sizes of sub-arrays, we evaluate the impact of the array aperture on the UE positioning performance. The sub-arrays are defined in Fig.~\ref{fig:UPA photo}. The RMSE and PEBs for different sub-arrays are illustrated in Fig.~\ref{fig:PEBvsArraySize}. It can be observed that when the SNR is high, the  maximum likelihood estimator achieves RMSEs comparable to those of the associated PEBs, which validates not only the effectiveness of the maximum likelihood estimator but also the correctness of our derived PEBs. The results also show that, with a fixed SNR, positioning with a larger array aperture contributes to a lower error bound, which is expected and consistent with the finding in~\cite{Z. Wang2025} that used an ideal array model. This can be explained by further decomposing the entries of $\mathbf{F}_{\Theta}$. Taking the example of $[\mathbf{F}_{\Theta}]_{\theta\theta}$ in~(\ref{eq:F_thetatheta}), it can be decoupled as 
\begin{equation}
\left[ \mathbf{F}_{\Theta_k} \right]_{\theta\theta}=\frac{2E_s}{\sigma^2} \mathcal{R}\left\{ \sum_{\text{p}\in\{V,H\}}\sum_{\text{p}'\in\{V,H\}}\sum_{n=1}^{N}  \Gamma_{n,\text{pp}'}\right\}, 
\label{eq:F_thetatheta2}
\end{equation}
where
$\Gamma_{n,\text{pp}'}=\big(\text{Vec}\big(\mathbf{Q}_{T_p}^{(n)}\big)^{\text{T}}\cdot \big[(\mathbf{\alpha}_{\theta}\odot\mathbf{w}_{\theta})\otimes\mathbf{w}_{\phi}\big]\big)\big( \alpha_{\text{pp}}   c_{R_{\text{p}}}'   \big)^*\big( \alpha_{\text{pp}}c_{R_{\text{p}}}'\big) .$
Clearly, a larger array size $N$ contributes to a larger FIM $\left[ \mathbf{F}_{\Theta_k} \right]_{\theta\theta}$. Similar conclusions can be drawn in terms of FIM $\left[ \mathbf{F}_{\Theta_k} \right]_{\phi\phi}$ and FIM $\left[ \mathbf{F}_{\Theta_k} \right]_{\tau\tau}$ by further decomposing~(\ref{eq:F_phiphi}) and~(\ref{eq:F_taotao}). Therefore, the higher FIM values for $\theta$, $\phi$, and $\tau$ contribute to a lower bound for the UE positioning error, as shown in~(\ref{eq:F_local}) and~(\ref{eq:PEB}). 
\subsubsection{Comparison with Ideal Array models} The PEBs achieved for an ideal array model are also shown in Fig.~\ref{fig:PEBvsArraySize}a and b for comparison. Here, the ideal array refers to one without any inter-element imbalances, and each element exhibits perfect XPI. Clearly, significant discrepancies in PEB performance are observed between EADF-based and ideal array models. Since the EADF-based model is entirely derived from measured radiation patterns, it inherently captures realistic array effects such as manufacturing imperfections, frame asymmetry, and mutual coupling.
The comparison results indicate that the performance impact of such practical effects cannot be ignored.
Hence, assuming an ideal array may lead to overly optimistic and unrealistic performance analyzes for UE positioning.   



\subsubsection{Impact of Spatial Distribution of APs}
\begin{figure}[t]
	\centering
		\begin{minipage}[tb]{0.45\textwidth}
			\centering
			\includegraphics[width=1\textwidth]{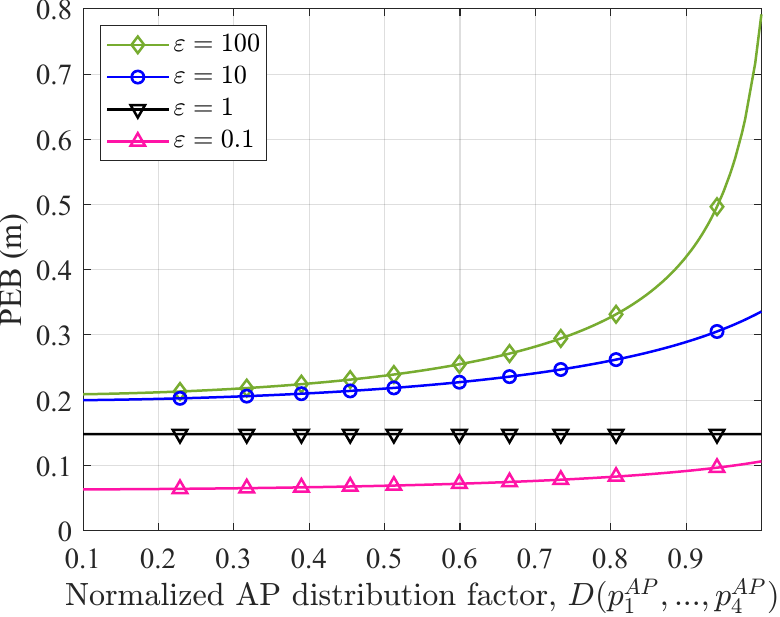}
		\end{minipage}
	\caption{PEBs with respect to the normalized AP geometry factor $D(\mathbf{p}_1^{\text{AP}},...,\mathbf{p}_4^{\text{AP}})$, where the weight factor $\epsilon$ is defined as $\epsilon=\frac{[\mathbf{F}_{\Theta_k}]_{\theta\theta} }{\tau^2[\mathbf{F}_{\Theta_k}]_{\tau\tau} }$.}
	\label{fig:PEBvsDistributionFactor}
\end{figure}

To investigate the relationship between the PEBs and the spatial distributions of APs, we first analyze the PEB with respect to the defined AP geometry factor $D\left(\mathbf{p}^{\text{AP}}_1,...,\mathbf{p}^{\text{AP}}_K\right)$ in Theorem~3. A 2D scenario with four distributed APs, each equipped with the described sub-6~GHz UPA, is considered. The AP positions are randomly generated, while the propagation distances between each AP and the UE are fixed at 5~m under different receive SNRs, i.e., 10, 15, 20, and 25~dB. The angle between each AP and the UE is assumed to remain unchanged. Fig.~\ref{fig:PEBvsDistributionFactor} shows the relationships between the associated PEB and $D\left(\mathbf{p}^{\text{AP}}_1,...,\mathbf{p}^{\text{AP}}_4\right)$. It can be observed that the PEB exhibits a non-decreasing trend with $D\left(\mathbf{p}^{\text{AP}}_1,...,\mathbf{p}^{\text{AP}}_4\right)$. When $\epsilon=\frac{[\mathbf{F}_{\Theta_k}]_{\theta\theta} }{\tau^2[\mathbf{F}_{\Theta_k}]_{\tau\tau} }=1$, the distance and azimuth information intensities of each AP are equal. In this case, the information ellipse becomes circular, as discussed earlier, indicating that the geometric distribution of the APs does not affect the limits of positioning performance. These results validate that the defined factor $D\left(\mathbf{p}^{\text{AP}}_1,...,\mathbf{p}^{\text{AP}}_K\right)$ can serve as a sufficient indicator to evaluate the effect of the spatial distribution of APs on the PEB. 

\begin{figure}[t]
	\centering
	\subfloat[]{
		\begin{minipage}[tb]{0.40\textwidth}
			\centering
			\includegraphics[width=1\textwidth]{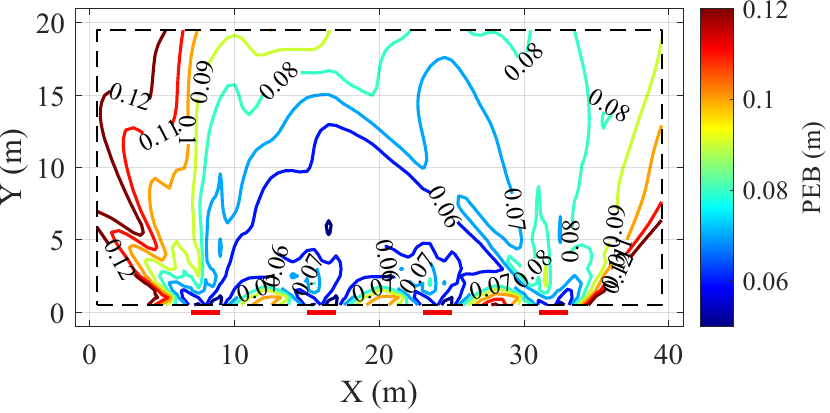}
		\end{minipage}
	}\vspace{-2mm}
	\\
	\subfloat[]{
		\begin{minipage}[tb]{0.40\textwidth}
			\centering
			\includegraphics[width=1\textwidth]{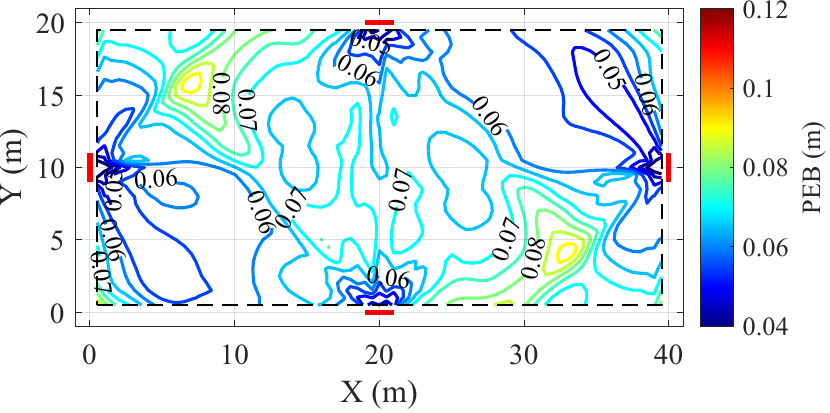}
		\end{minipage}
	}\vspace{-2mm}
	\\
	\subfloat[]{
		\begin{minipage}[tb]{0.40\textwidth}
			\centering
			\includegraphics[width=1\textwidth]{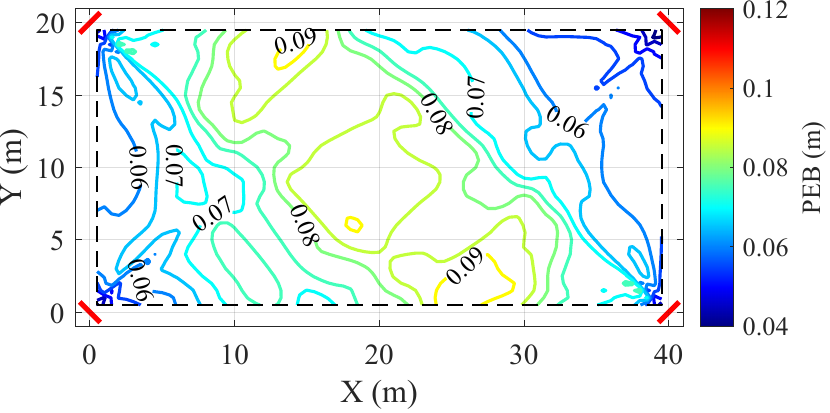}
		\end{minipage}
	}
	\caption{PEB contours for different AP spatial distributions: (a) APs evenly distributed along one side of the environment, (b) APs located in the middle of each side in the environment, and (c) APs deployed in the four corners of the environment. }
	\label{fig:PEB contours}
\end{figure}

Next, we analyze some typical AP deployment scenarios and compare their corresponding PEB performance. Specifically, we consider four APs deployed in an indoor environment with the following configurations: 1) APs evenly placed along one side of the environment; 2) APs located in the middle of each side in the environment; 3) APs deployed at the four corners of the environment. The PEB contours for these three cases are illustrated in Fig.~\ref{fig:PEB contours}.
In Fig.~\ref{fig:PEB contours}a, the PEB is the lowest near the APs and increases rapidly with distance. Because all APs are placed on the same side, the available positioning information is confined to a narrow azimuthal range. Consequently, UEs located farther from the APs experience larger positioning errors, especially at the corners on the same side, indicating non-uniform coverage. In Fig.~\ref{fig:PEB contours}b and Fig.~\ref{fig:PEB contours}c, where the APs are distributed in the middle of each of the four sides and in the four corners, the PEBs become more uniform throughout the area, with lower maximum errors than in the first case. These layouts improve coverage and positioning accuracy for UEs throughout the whole environment, while also minimizing the worst-case positioning error and maintaining robust localization performance at arbitrary UE locations. The results highlight the significant impact of the spatial distribution of APs on the positioning performance limits. Deploying APs to experience more spatial diversity leads to a more consistent and improved positioning performance in the target region. 

\subsubsection{Impact of UE Tilt}
\begin{figure}[t]
	\centering
		\begin{minipage}[tb]{0.45\textwidth}
			\centering
			\includegraphics[width=1\textwidth]{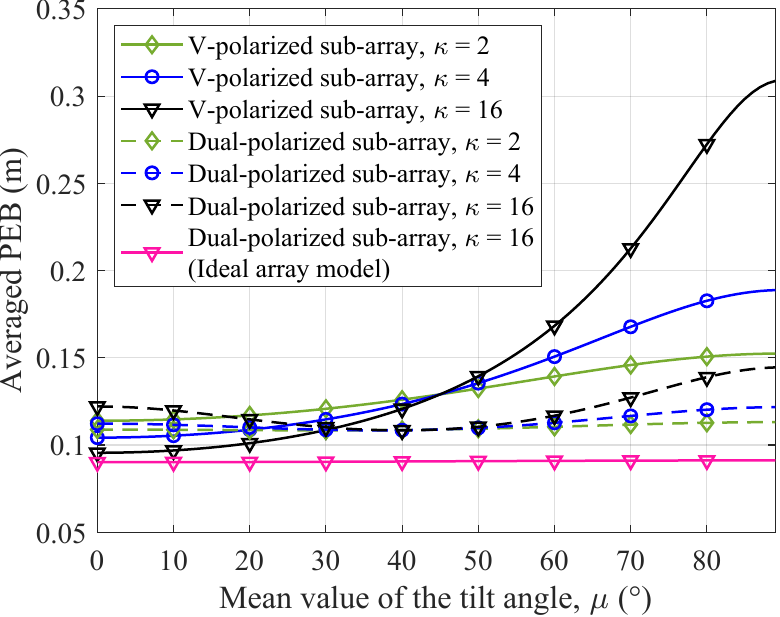}
		\end{minipage}
	\caption{Averaged PEBs with respect to the tilt angle $\beta$ for the single- and dual-polarized sub-arrays.}
	\label{fig:PEBvsBeta}
\end{figure}

In the following experiments, two sub-arrays from the sub-6~GHz UPA are selected as the AP-side antennas. The first is a single-polarized array comprising only vertical polarization elements $\{1V,...,8V\}$, while the second is a dual-polarized array composed of alternating vertical and horizontal elements $\{1V,2H,3V,4H,5V,6H,7V,8H\}$. 

Given different $\mu$ and $\kappa$, the averaged PEBs are evaluated through Theorem~4. The results are illustrated in Fig.~\ref{fig:PEBvsBeta}. 
We recall that as $\kappa$ increases, the distribution of $\beta$ becomes more concentrated around the mean $\mu$. For the single-polarized array, the averaged PEB increases as $\mu$ increases, indicating a degradation in positioning accuracy. This is because as $\mu$ approaches $\pi/2$, corresponding to a nearly horizontal UE orientation, the V-polarized array experiences an increased polarization mismatch, significantly reducing the received signal power and, thereby, decreasing the positioning accuracy. In contrast, the dual-polarized array exhibits much more stable PEBs across varying values of $\mu$ and $\kappa$. This robustness comes from its ability to capture both V- and H-polarized signals, thereby mitigating the adverse effects of the UE tilt. Consequently, the dual-polarized setup enables more consistent and reliable positioning performance regardless of the UE tilt.

For comparison, the case of the dual-polarized setup based on the ideal array model is also analyzed. The resulting PEB remains constant for $\mu$, as shown in Fig.~\ref{fig:PEBvsBeta}. This behavior is expected since the model assumes ideal elements with perfect XPI, thereby capturing both V- and H-polarized signals equally efficiently. In contrast, in the same configuration, the PEB obtained with the EADF-based model varies with $\mu$. These variations are typically caused by practical array effects, namely, intra- and inter-element imbalances resulting from manufacturing imperfections, frame asymmetry, and mutual coupling. Such imbalances are ignored by the ideal model but are inherently captured by the EADF that is entirely derived from measured radiation patterns. The results highlight the performance gap between idealized and realistic antenna models and indicate the importance of incorporating practical antenna properties in positioning analysis.   

\vspace{-2mm}
\section{3D Cooperative UE Tracking in Distributed MIMO Systems}
\begin{figure}[t]
	\centering
		\begin{minipage}[tb]{0.48\textwidth}
			\centering
			\includegraphics[width=1\textwidth]{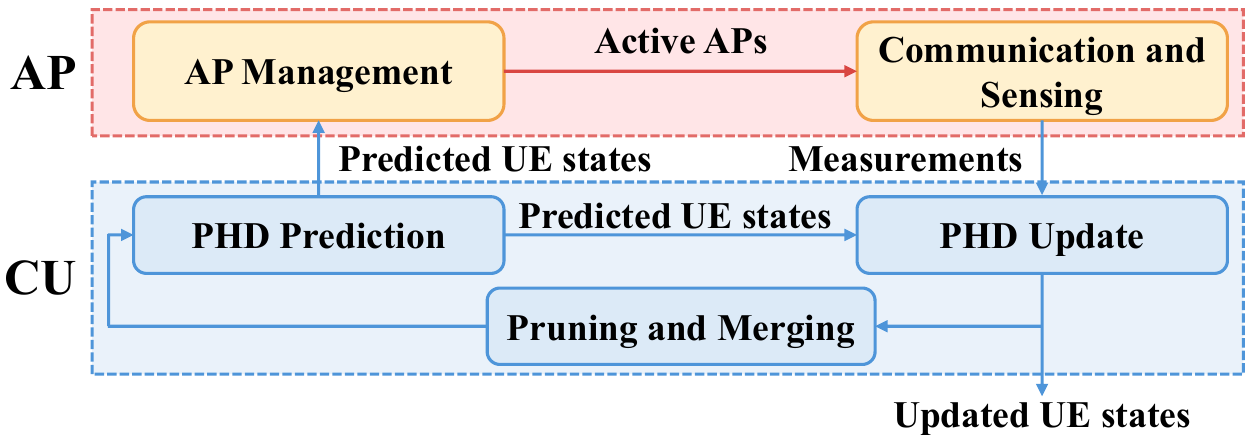}
		\end{minipage}
	\caption{Workflow of the complete 3D cooperative UE tracking framework.}
	\label{fig:workflow}
\end{figure}

This section introduces a comprehensive UE tracking framework for distributed MIMO systems. The goal is to continuously track mobile UEs based on the MPC parameters, referred to as the \emph{measurements} for UE tracking in the subsequent analysis, which are extracted from the received signals. All APs are assumed to be connected to a central/control unit (CU), which processes the measurements from the APs. At each time step, a subset of APs may be active for communication with the UEs and for sharing their measurements with the CU. The activation of each AP is determined according to a predefined AP management strategy. Based on the fused measurements, a global PHD filter is performed at the CU to enable continuous UE tracking. An overview of the proposed framework is illustrated in Fig.~\ref{fig:workflow}. In the following subsections, we first present the state and measurement models, followed by the implementation of the global PHD filter. Furthermore, based on the PEB analyzed in Section~III, we propose a PEB-aware AP management strategy to optimize AP activation for energy-efficient and accurate UE tracking.

\subsection{State and Measurement Models}
\subsubsection{State Model}
The state of the $k$-th AP includes its 3D location $\mathbf{p}^{\text{AP}}_k$ and 3D orientation $\mathbf{r}_k^{\text{AP}}$, which are considered \emph{a priori}. The state $\mathbf{x}_{m}^t$ of the $m$-th mobile UE at time step $t$ is characterized by its 3D position $\mathbf{p}^{\text{UE}}_{m}(t)$ and velocity $\mathbf{\upsilon}_{m}^t$. The state transition density is expressed as 
\begin{equation}
f_{t+1|t}\left(\mathbf{x}_{m}^{t+1}|\mathbf{x}_{m}^t\right)=\mathcal{N}\left(\mathbf{x}_{m}^{t+1};\zeta \left(\mathbf{x}_{m}^t\right),\mathbf{Q}^{t+1}\right),
\label{eq:transition model}
\end{equation}
where $\zeta(\cdot)$ and $\mathbf{Q}$ denote the transition model (e.g., a constant-velocity model) and the process noise covariance, respectively.  

\subsubsection{Measurement Model}
To solve the UE tracking problem, we consider a two-stage process. First, after receiving the signals, channel estimation is performed at each AP, and the estimated MPC delays and AoAs are obtained. Second, assuming that the measurements of different APs are independent and homogeneous, the delays and AoAs are transformed into \emph{position-related measurements} $\tilde{\mathbf{Z}}^t=\{ \tilde{\mathbf{z}}_{1,k}^t,...,\tilde{\mathbf{z}}_{L,k}^t\}_{k=1}^{K}$. This is achieved using the inverse sigma points~\cite{A. F. Garcia-Fernandez2015}, which is a common practice to linearize non-linear measurements. Then the clustering method in~\cite{T. Li2017} is adopted to group those multi-source position-related measurements corresponding to the same objects and generate a `proxy' measurement. Let $\mathbf{Z}^t=\{ \mathbf{z}_{1}^t,...,\mathbf{z}_{L'}^t\}$ denote the set of these `proxy' measurements which are used in the PHD update.

The measurement model for the UE tracking is expressed as
$\mathbf{z}_{l,m}^t = \mathbf{p}^{\text{UE}}_{m}(t) + \mathbf{n}^{t}_{l,m}$,
where $\mathbf{n}^{t}_{l,m}$ denotes the measurement noise with a covariance of $\mathbf{W}^{t}_{l,m}$. The likelihood function is thus expressed as
\begin{equation}
h\left(\mathbf{z}_{l,m}^t|\mathbf{x}_{m}^t\right)=\mathcal{N}\left(\mathbf{z}_{l,m}^t;\xi\left(\mathbf{x}_{m}^t\right),\mathbf{W}^{t}_{l,m}\right),
\label{eq: measurement model}
\end{equation}
with $\xi\left(\mathbf{x}_{m}^t\right)=\mathbf{p}^{\text{UE}}_{m}(t)$.
Note that there are non-LoS (NLoS) paths in the received measurements, i.e., MPCs that may have experienced reflection, scattering, or diffraction. In this work, such paths are treated as clutter\footnote{While there are methods that utilize NLoS paths for UE tracking, as discussed in~\cite{H. Kim2020_1,O. Kaltiokallio2024,H. Kim2020_2,Y. Ge2024}, such approaches are beyond the scope of this work.}. Furthermore, it is unclear which measurements originate directly from the UEs or from clutter, making it difficult to perform an accurate data association. To manage this uncertainty, 
the set of UE states at time $t$ is modeled as an RFS $\mathbf{X}^t=\left\{ \mathbf{x}_{1}^t,...,\mathbf{x}_{M}^t\right\}$. Since LoS paths may also be undetected in some time steps, a detection probability $p_{\text{D}} \in [0, 1]$ is introduced to quantify the probability that a given measurement comes directly from a UE. Based on this model, we employ a PHD filter, where the measurement sets $\mathbf{Z}^t$ are used to estimate the UE state RFS $\mathbf{X}^t$, as will be detailed in the following section. 

\subsection{Global PHD Filter for UE Tracking}
Based on a Bayesian filtering framework, a PHD filter aims to recursively estimate the first-order statistical moment of the posterior density of the RFS $p(\mathbf{X}^t|\mathbf{Z}^t)$, referred to as the PHD $v_t(\mathbf{x})$. The local maxima of $v_t(\mathbf{x})$ provide estimates of the RFS $\mathbf{X}^t$ of the UE state, allowing for the extraction of individual UE states. Specifically, the PHD propagates over time through the following prediction and update steps:
\begin{equation}
\begin{split}
v_{t|t-1}(\mathbf{x}) = \int f_{t|t-1}(\mathbf{x}|\mathbf{x}^{t-1}) v_{t-1}(\mathbf{x}^{t-1}) \text{d} \mathbf{x},
\label{eq:PHD rescure1}
\end{split}
\end{equation}
\begin{equation}
\begin{split}
    v_t(\mathbf{x}) =& (1 - p_D) v_{t|t-1}(\mathbf{x}) \\
    & + \sum_{\mathbf{z} \in \mathbf{Z}^t} \frac{p_D h_t(\mathbf{z}|\xi(\mathbf{x})) v_{t|t-1}(\mathbf{x})}{\lambda_c + p_D \int h_t(\mathbf{z}| \xi(\mathbf{x})) v_{t|t-1}(\mathbf{x}) \text{d}\mathbf{x}},
    \label{eq:PHD rescure2}
\end{split}
\end{equation}
where $f(\cdot)$, $h(\cdot)$, and $\lambda_c$ denote the state transition density, the likelihood function, and the clutter intensity, respectively. To obtain a tractable solution, the PHD is modeled as a Gaussian mixture (GM)
	$v_t(\mathbf{x}) = \sum_{i=1}^{J(t)} w_i^t \mathcal{N}\left(\mathbf{x}; \mathbf{m}_i^t, \mathbf{P}_i^t\right)$~\cite{K. Granstrom2012_1},
with $J(t)$, $w^t$, $\mathbf{m}^t$, and $\mathbf{P}^t$ denoting the number of GM components, the component weight, the component mean, and the component covariance at time $t$, respectively. Accordingly, the prediction step~(\ref{eq:PHD rescure1}) can be rewritten as
\begin{equation}
	v_{t|t-1}(\mathbf{x}) = \sum_{i=1}^{J_{t|t-1}} w_i^{t|t-1} \mathcal{N}\left(\mathbf{x}; \mathbf{m}_i^{t|t-1}, \mathbf{P}_i^{t|t-1}\right),
	\label{eq:GM-PHD1}
\end{equation}
with $w_i^{t|t-1}=w_i^{t-1}$, $\mathbf{m}_i^{t|t-1} = \zeta(\mathbf{m}_i^{t-1})$, and $\mathbf{P}_i^{t|t-1} = \zeta\left(\mathbf{m}_i^{t-1}\right)\mathbf{P}_i^{t-1} \left[\zeta\left(\mathbf{m}_i^{t-1}\right)\right]^{\text{T}}+\mathbf{Q}^{t-1}$.
The update step~(\ref{eq:PHD rescure2}) is given by
\begin{equation}
\begin{split}
	v_t(\mathbf{x}) &= (1 - p_D) v_{t|t-1}(\mathbf{x})+\sum_{\mathbf{z} \in \mathcal{Z}^t}\sum_{i=1}^{J_{t|t-1}}w^{t|t}_i\mathcal{N}\left(\mathbf{x};\mathbf{m}_{i}^{t|t},\mathbf{P}_{i}^{t|t}\right),
	\label{eq:GM-PHD2}
\end{split}
\end{equation}
where
\begin{equation}
\begin{split}
	 w^{t|t}_i = \frac{p_Dw^{t|t-1}_i\mathcal{N}\left(\mathbf{z};\xi\left(\mathbf{m}^{t|t-1}_{i}\right),\mathbf{S}_{i}^{t}\right)}{\lambda_c + p_D \sum_{i=1}^{J_{t|t-1}}w^{t|t-1}_i\mathcal{N}\left(\mathbf{z};\xi\left(\mathbf{m}^{t|t-1}_{i}\right),\mathbf{S}_{i}^{t}\right)},
\end{split}
\end{equation}
\begin{equation}
\mathbf{m}_i^{t|t}=\mathbf{m}_i^{t|t-1}+\mathbf{K}_i^{t}\left(\mathbf{z}^{t}-\xi\left(\mathbf{m}_i^{t|t-1}\right)\right),
\end{equation}
\begin{equation}
\mathbf{P}_i^{t|t}=\mathbf{P}_i^{t|t-1}-\mathbf{K}_i^t\mathbf{S}_i^t\left[\mathbf{K}_i^t\right]^\text{T},
\end{equation}
\begin{equation}
\mathbf{K}_i^{t}=\mathbf{P}_i^{t|t-1}\xi\left(\mathbf{m}_i^{t|t-1}\right)^\text{T}\left[\mathbf{S}_i^t\right]^{-1},
\end{equation}
\begin{equation}
\mathbf{S}_i^{t}=\xi\left(\mathbf{m}_i^{t|t-1}\right)\mathbf{P}_i^{t|t-1}\xi
\left(\mathbf{m}_i^{t|t-1}\right)^\text{T}+\mathbf{W}^t.
\label{eq:16}
\end{equation}
After the update, the estimated UE-state RFS $\hat{\mathbf{X}}^t$ is extracted by selecting the $M$ Gaussian components with the largest weights,
$\hat{\mathbf{X}}^t=\big\{\mathbf{m}_i^{t|t}:i\in \mathbf{I}=\arg_{\{ i_1,...,i_M\}}w_{i_1}^{t|t}\geq w_{i_2}^{t|t}\geq...\geq w_{i_{J_{t}}}^{t|t}\big\}.$
To avoid exponential growth of Gaussian components over time, pruning and merging techniques are applied after each update~\cite{B.-N. Vo2006}. Components with negligible weights are discarded and components with similar parameters are merged into a single one.

We recall that a key feature of the introduced global filter lies in its multi-source measurement transformation before each PHD update. That is, AoA and delay measurements from different APs are first transformed into position-related measurements. Then, those multi-source position-domain measurements corresponding to the same objects are represented by a `proxy' measurement. This differs from existing filters~\cite{H. Kim2020_1,O. Kaltiokallio2024,H. Kim2020_2}, in which AoA and delay measurements are used directly in the nonlinear measurement model, leading to the need to approximate them through a nonlinear filter, which generally exhibits high computational complexity.

\subsection{PEB-Aware AP Management}
An effective strategy to determine which APs are active at each time step is essential to maintain continuous UE tracking while minimizing system overhead. Existing antenna selection strategies for UE tracking typically fall into two categories: FoV-based and error bound-based selection. The FoV-based approach dynamically selects APs that can cover current UE locations~\cite{Y. Xu2025}. 
This method is computationally efficient; however, it faces practical limitations, especially when APs are equipped with realistic antenna arrays, which makes accurate FoV modeling difficult. The error bound-based approach selects APs that are expected to provide a lower error bound for tracking. In this context, performance bounds such as the posterior CRLB~\cite{R Tharmarasa2007} and the Bayesian CRLB~\cite{J.Sun2023} are widely used as management criteria. However, these bounds cannot be directly computed if the explicit data association between measurements and targets or clutter is unknown, which is a key feature of the PHD filter. In this work, we use the PEB derived in Section~III. Since the PEB reflects how accurately a UE can be localized given its relative geometry to the APs, it inherently indicates the maximum achievable quality of the measurements. Therefore, APs associated with lower PEBs are more likely to contribute to high-quality measurements and ultimately to improved tracking performance.

The dynamic AP management problem is to determine the activation status of each AP at each time step. Specifically, it involves selecting a time-varying subset $\Gamma_t$ consisting of $K'$ APs that are responsible for communication, channel estimation, and measurement sharing with the CU. Assume that the predicted PHD at time $t+1$ has been computed using~(\ref{eq:GM-PHD1}), and the predicted UE states are obtained as $\hat{\mathbf{X}}^{t+1|t}=\{\mathbf{m}_i^{t+1|t}:i\in \mathbf{I}=\arg_{\{ i_1,...,i_M\}}w_{i_1}^{t+1|t}\geq w_{i_2}^{t+1|t}\geq...\geq w_{i_{J_{t}}}^{t+1|t}\}$, which include the predicted UE positions $\mathbf{p}_{m}^{\text{UE}}(t+1|t)$. We aim to optimize the AP activation indicator $\mathbf{g}_{t+1,k} \in \{ 0,1\},k=1,...,K$, such that 
\begin{equation}
\begin{split}
\mathbf{g}_{t+1}^{\text{Optimal}}=&\arg \min_{\mathbf{g}_{t+1}} \sum_{m=1}^{M}\mathcal{P}\left(\mathbf{p}_m^{\text{UE}}(t+1|t)\right)\\
=&\arg \min_{\mathbf{g}_{t+1}} \sum_{m=1}^{M}\sqrt{\text{tr}\left\{ \left[\sum_{k=1}^{K} g_{t+1,k}\mathbf{F}_{\mathbf{p},k} \right]^{-1} \right\}} 
\end{split}
\label{eq:g_optimal}
\end{equation}
subject to 
$\sum_{k=1}^{K}g_{t+1,k}=K'$.
Here $g_{t+1,k}=1$ indicates that AP $k$ is active at time $t+1$; otherwise, it is inactive. The selected AP set is then given by $\Gamma_t = \{k|g_{t+1,k}=1\}$. 

The optimization problem in~(\ref{eq:g_optimal}) is NP-hard. A brute-force search method is computationally feasible only when $K$ is small, but it quickly becomes intractable in dense AP deployments. To address this, we adopt an iterative greedy-local-search algorithm inspired by~\cite{R. Tharmarasa2007} to find a near-optimal solution. The algorithm contains two main steps: 1) \emph{Greedy initialization}, which starts with the AP that individually provides the smallest PEB (i.e., the most informative AP). Then iteratively add one AP at a time that, when included, reduces the overall PEB the most. This continues until $K'$ APs are selected. The selected APs form the set $\Gamma_{t+1}$; 2) \emph{Local search refinement}, which iteratively evaluates the replacements of a selected AP with one not currently in $\Gamma_{t+1}$. If any such swap leads to a lower total PEB, update $\Gamma_{t+1}$. This repeats until no further improvement is possible. The complete procedure is summarized in Algorithm~\ref{algorithm1}. 

\begin{algorithm}[tb!]
	\renewcommand{\algorithmicrequire}{\textbf{Input:}}
	\renewcommand{\algorithmicensure}{\textbf{Output:}}
\caption{An iterative greedy-local-search method to solve the optimization problem in~(\ref{eq:g_optimal}). }
\label{algorithm1}
\begin{algorithmic}[1]
\STATE Initialize AP index set $\Upsilon=\{1,...,K\}$, AP subset $\Gamma_t=\varnothing$, $\tilde{K}'=0$ 
\WHILE{$\tilde{K}'\leq K'$}
\STATE $\tilde{K}'\leftarrow\tilde{K}'+1$ 
\STATE Solve~(\ref{eq:g_optimal}) given that $g_{t,k}=1,k\in\Gamma_t$, and get $\mathbf{g}_{t}^{\text{Optimal}}$.
\STATE $\Gamma_t \leftarrow \{k\in\Upsilon|g_{t,k}^{\text{Optimal}}=1\}$.
\ENDWHILE
\STATE Define $\bar{\Upsilon}\leftarrow\Upsilon-\Gamma_t $ and $D=\sum_{m=1}^{M}\mathcal{P}\left(\mathbf{p}_m^{\text{UE}}(t|t-1)\right)$ given $\mathbf{g}_{t}|_{\Gamma_t}$.
\WHILE {$D$ keeps decreasing}
\FOR{$k_1\in \bar{\Upsilon},k_2\in \Gamma_t$}
\STATE $\Gamma_t'\leftarrow\left\{\Gamma_t-\{k_2\}\right\}+\{k_1\}$ 
\STATE Calculate $D'=\sum_{m=1}^{M}\mathcal{P}\left(\mathbf{p}_m^{\text{UE}}(t|t-1)\right)$ given $\mathbf{g}_{t}|_{\Gamma_t'}$.
\IF{$D'<D$}
\STATE $D\leftarrow D$, $\Gamma_t\leftarrow \Gamma_t'$, and $\bar{\Upsilon}\leftarrow\Upsilon-\Gamma_t $.
\ENDIF
\ENDFOR
\ENDWHILE
\STATE Return $\Gamma_t$ and $\mathbf{g}_{t}|_{\Gamma_t}$.
\end{algorithmic}			
\end{algorithm}

Once $\Gamma_{t+1}$ is determined, the responsibilities for communication, channel estimation, and measurement sharing to the CU are transferred from the APs in $\Gamma_{t}$ to those of $\Gamma_{t+1}$. The subsequent update of the UE state (\ref{eq:PHD rescure2}) is then performed with the collected measurements $\mathcal{Z}^{t+1}=\{\mathbf{Z}_k^{t+1}:k\in \Gamma_{t+1} \}$.

\section{Tracking Performance Evaluation with Real-World Channels}
In this section, the proposed UE tracking framework is evaluated using practical channels. The measurement environment and setup are first described, followed by an evaluation of the proposed framework for the measured channels. 
\subsection{Indoor Distributed MIMO Channel Measurements}
\begin{figure}[t]
	\centering
	\begin{minipage}[tb]{0.35\textwidth}
		\centering
		\includegraphics[width=1\textwidth]{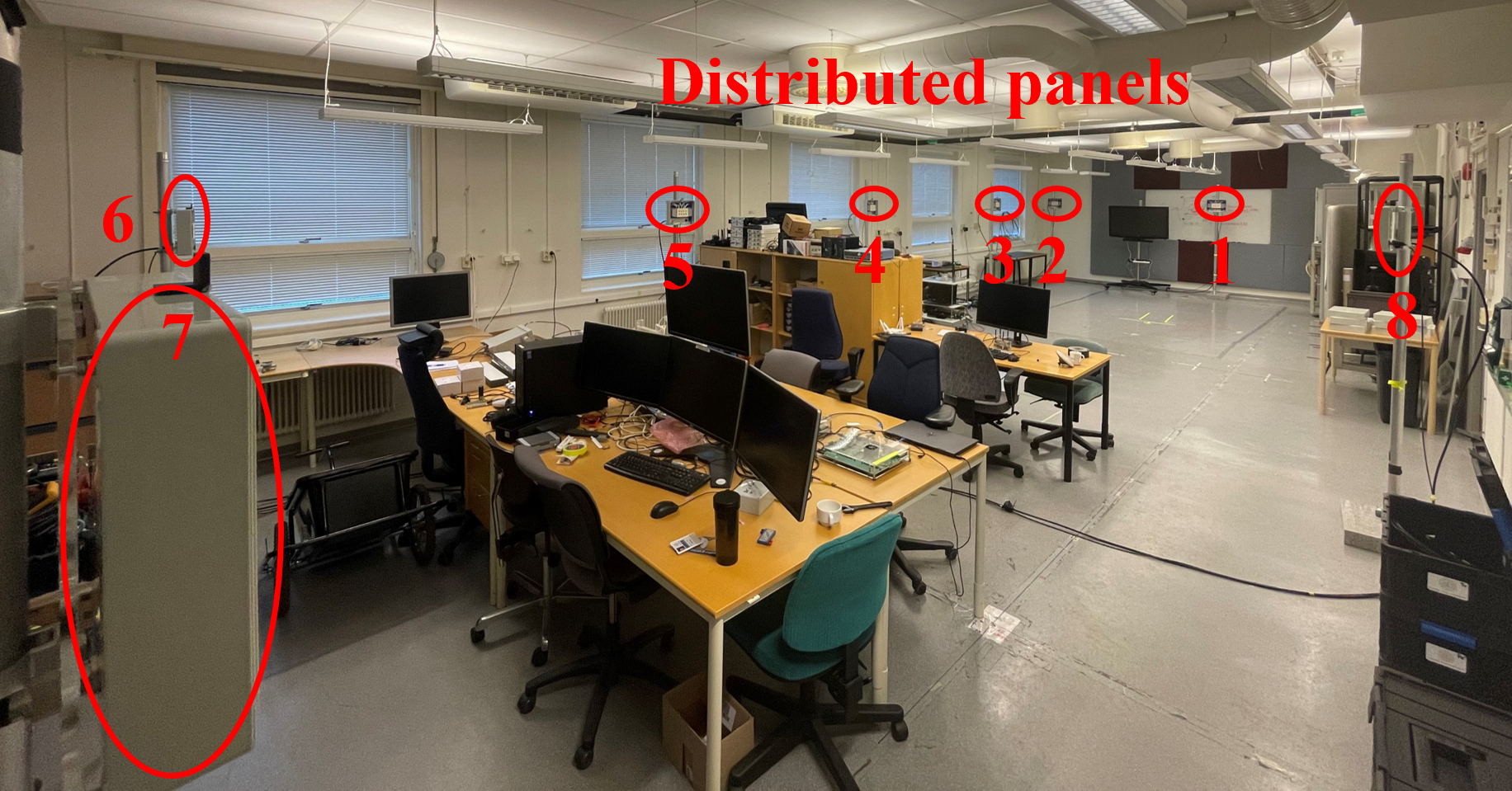}
	\end{minipage}
	\caption{Photo of the measurement environment.}
	\label{fig:Measurement_environments}
\end{figure}

The measurement campaign was carried out in an indoor lab, as illustrated in Fig.\ref{fig:Measurement_environments}(a). A wideband switched-array channel sounder\cite{M. Sandra2024} was used for channel collection. On the AP side, eight sub-6~GHz UPAs (see Fig.\ref{fig:UPA photo}a), referred to as `panels', were deployed. On the UE side, a single omnidirectional antenna was used. Before measurements were performed, a back-to-back calibration was conducted to eliminate the influence of the sounder hardware, connectors, and cables.

The measurements were performed at a carrier frequency of 5.6~GHz with a bandwidth of 400~MHz. The panels were distributed throughout the room, as shown in Fig.~\ref{fig:Measurement_environments}. The UE antenna was vertically mounted on a mobile robot, which moved from one side of the room to the other. A global coordinate system was established with its origin set near the UE's starting location. The ground truth of the UE trajectory was accurately recorded using a Lidar sensor. In total, uplink channel measurements were collected for 780 discrete UE positions (corresponding to 780 time steps). The MPC parameters, including delay, AoA, Doppler frequency, and polarization matrix, were extracted from the measured channels using the SAGE algorithm~\cite{X. Yin2003}. Further details on the algorithm configuration can be found in~\cite{Y. Xu2025,M. Sandra2024}.

\subsection{Results and Analysis}
Using the measured channels, UE tracking was performed using the proposed framework. A random walk model~\cite{E. Rastorgueva-Foi2024} was adopted in~(\ref{eq:transition model}), with the process noise covariance set to $\mathbf{Q}=\text{diag}[0.1^2 \text{ m}^2,0.1^2 \text{ m}^2,0.1^2 \text{ m}^2]$. The measurement noise covariance $\mathbf{W}$ in~(\ref{eq: measurement model}) for each MPC at each time was formulated based on the estimation covariance of the SAGE algorithm, as detailed in~\cite{Y. Xu2025}. The maximum number of Gaussian components was set at 500, with the pruning and merging threshold set at $10^{-4}$ and 4, respectively. To evaluate UE tracking performance, instantaneous RMSE was defined as $\eta(t)=\sqrt{\|\hat{\mathbf{p}}^{\text{UE}}(t)-\mathbf{p}^{\text{UE}}(t) \|^2}$.

\begin{figure}[t]
	\centering
	\begin{minipage}[tb]{0.43\textwidth}
		\centering
		\includegraphics[width=1\textwidth]{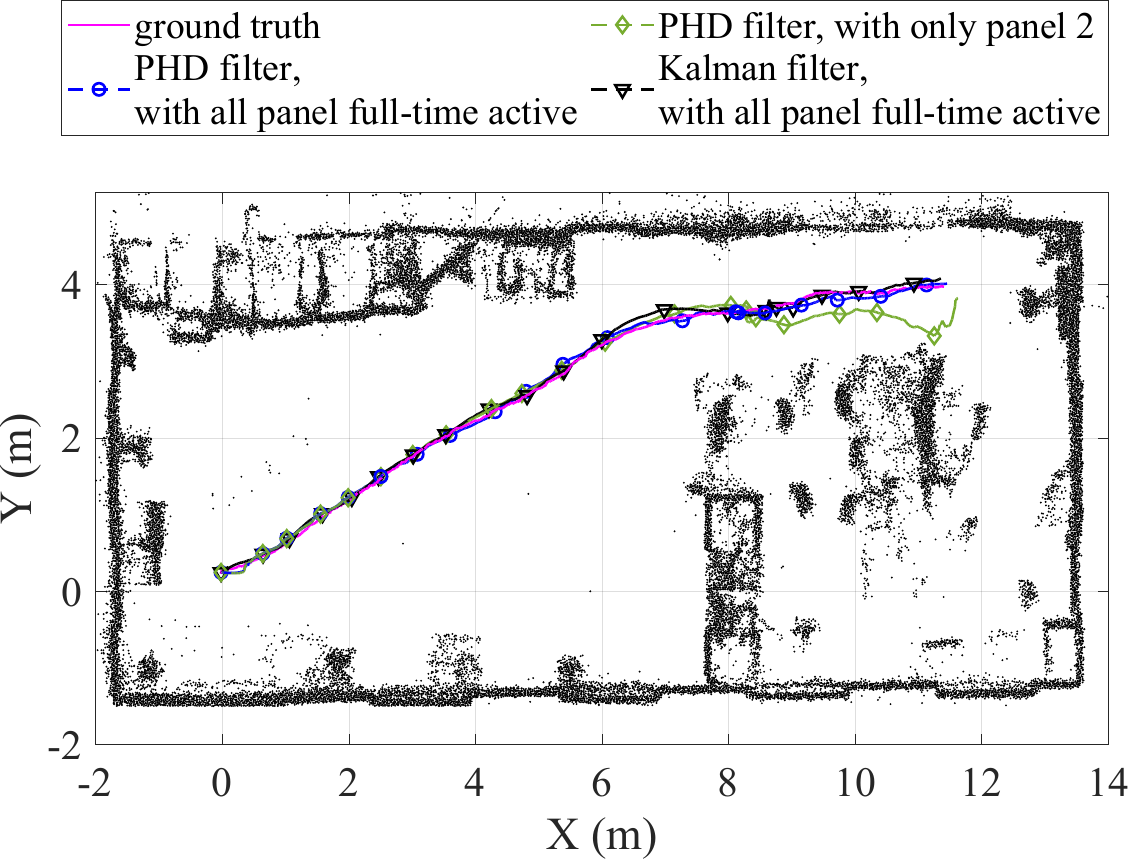}
	\end{minipage}
	\caption{Comparison of the ground truth and the tracked trajectories using the introduced PHD filter and a Kalman filter.}
	\label{fig:trajectories}
\end{figure}

\begin{figure}[t]
	\centering
	\begin{minipage}[tb]{0.45\textwidth}
		\centering
		\includegraphics[width=1\textwidth]{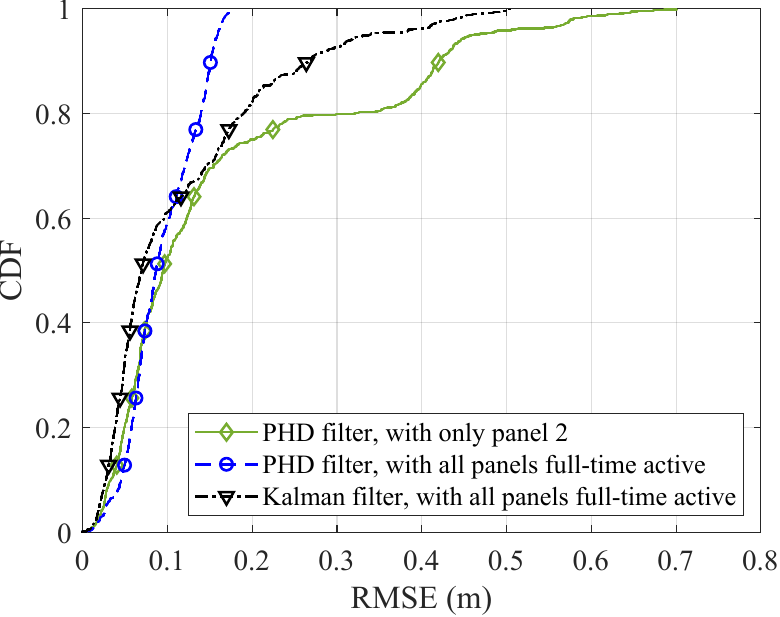}
	\end{minipage}
	\caption{CDFs of tracking RMSEs: proposed PHD filter (single panel vs. all panels active) and Kalman filter (all panels active).}
	\label{fig:RMSE_cdf}
\end{figure}

We first compare the tracking performance with two configurations: using only a single panel and using all panels. Panel~2 in Fig.~\ref{fig:Measurement_environments} is selected as a representative example for the single-panel case. The tracked trajectories for both configurations are shown in Fig.~\ref{fig:trajectories}, and the CDFs of their corresponding RMSEs are shown in Fig.~\ref{fig:RMSE_cdf}. The results show that single-panel tracking (panel~2) performs well when the UE is near, but it degrades significantly as the UE moves farther away, primarily due to reduced SNR. In contrast, tracking with all panels provides robust performance throughout the entire trajectory, owing to the more uniform spatial coverage offered by the distributed panels. The resulting average RMSE is 0.09~m. For comparison, a conventional Kalman filter~(KF)-based tracking method was also implemented. In the KF approach, data association is carried out using the nearest-neighbor principle, which represents a hard-decision association strategy. The KF was configured with the same parameters as the proposed PHD filter. The corresponding tracked trajectory and the CDF of the RMSE are presented in Figs.~\ref{fig:trajectories} and~\ref{fig:RMSE_cdf}, respectively. The results indicate that the KF-based method exhibits a inferior tracking performance, with an average RMSE of 0.12~m. This performance degradation is mainly due to incorrect data association, particularly in scenarios where the LoS path is blocked or not detected. Such issues are common in practice due to environmental obstructions, signal instability, or channel estimation errors. In contrast, the proposed PHD filter treats all measurements as an RFS, thus avoiding explicit data association and demonstrating superior robustness under realistic conditions.

\begin{figure}[t]
	\centering
	\begin{minipage}[tb]{0.45\textwidth}
		\centering
		\includegraphics[width=1\textwidth]{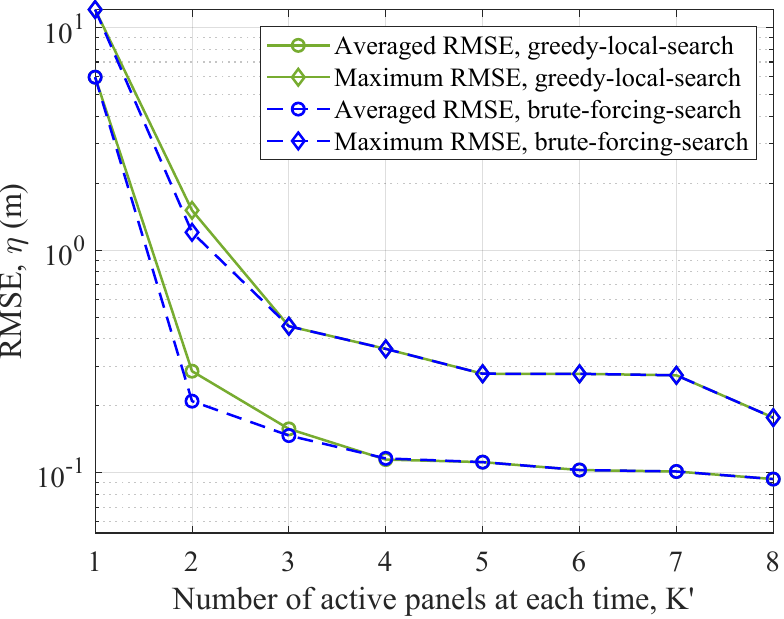}
	\end{minipage}
	\caption{Average and maximum RMSE with respect to the number of active panels at each time, $K'$ .}
	\label{fig:RMSEvsK}
\end{figure}

\begin{figure}[t]
	\centering
	\subfloat[]
	{
		\begin{minipage}[tb]{0.48\textwidth}
			\centering
			\includegraphics[width=1\textwidth]{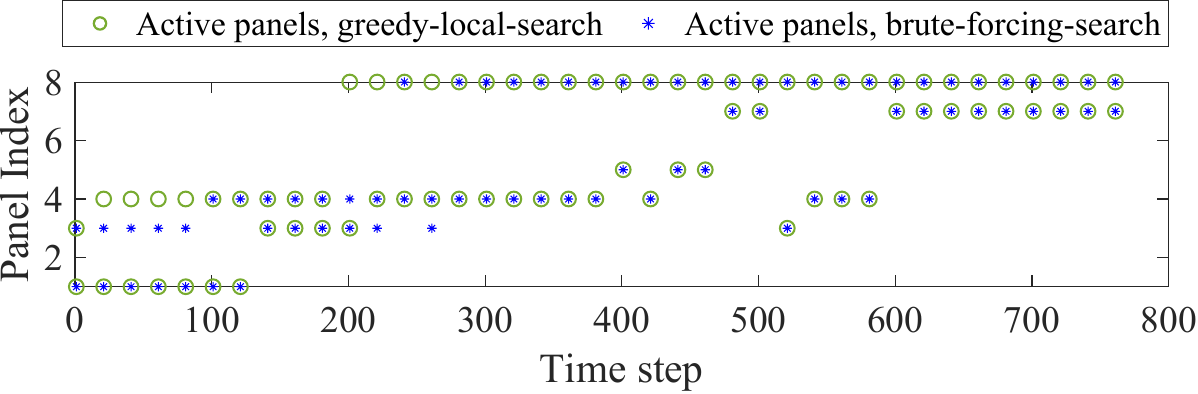}
		\end{minipage}
	}\vspace{-2mm}\\
	\subfloat[]
	{
		\begin{minipage}[tb]{0.48\textwidth}
			\centering
			\includegraphics[width=1\textwidth]{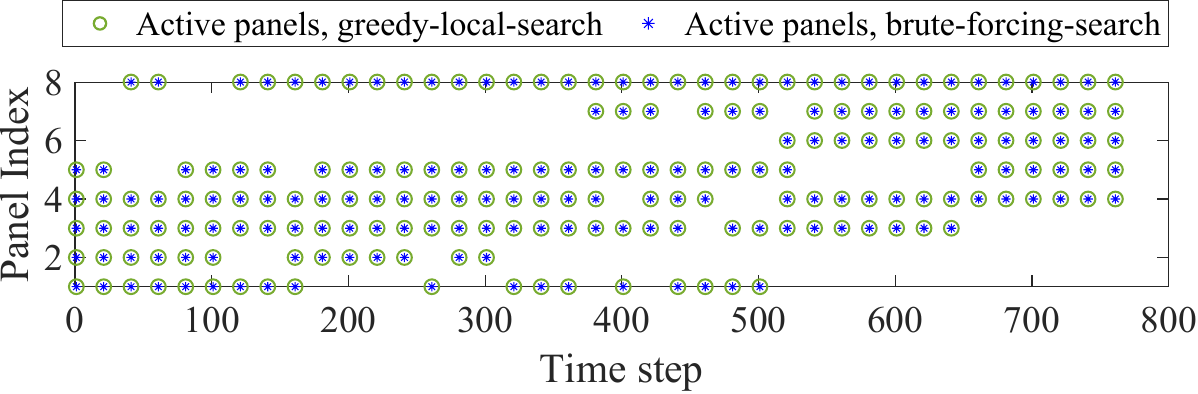}
		\end{minipage}
	}
	\caption{Activation status of the panels over time with (a) $K'=2$ and (b) $K'=5$.}
	\label{fig:ActivePanels}
\end{figure}

Furthermore, we evaluate the effect of AP management on the tracking. Different values of $K'$ are studied. The activation status of the panels is determined by the proposed PEB-aware AP management strategy, where a near-optimal solution is obtained using the iterative greedy-local-search method introduced in solving~(\ref{eq:g_optimal}). Since the total number of panels is relatively small in our measurements, the global-optimal solution is also obtained via a brute-force search for comparison. Fig.~\ref{fig:RMSEvsK} presents the maximum and average RMSEs corresponding to different $K'$. For a fixed $K'$, the greedy-local-search method yields a performance similar to that of the brute-force search. From $K'=5$ onward, both methods achieve identical RMSEs, as confirmed by the identical panel activation patterns shown in Fig.~\ref{fig:ActivePanels}. This result indicates that the greedy-local-search method successfully identifies the optimal solution in this scenario, further validating its robustness for AP management.

In addition, both the results with near-optimal and global-optimal AP management show that when $K'$ is small, the RMSEs decrease significantly with an increase of $K'$, indicating improved tracking performance that benefits from more panels working simultaneously. However, such improvement becomes more limited when $K'$ increases beyond a sufficient number. Specifically, when $K'=5$, the resulting average RMSE with global-optimal AP management decreases to around 0.10~m, which is quite close to that when all panels are fully active (i.e. $K'=8$), indicating that with the proposed AP management strategy, a smaller number of concurrent panels can achieve tracking performance similar to the all-active approach. This further demonstrates the effectiveness of the proposed method, which enables continuous and accurate UE tracking while reducing system overhead.          

\vspace{-2mm}
\section{Conclusions}

In this paper, we have thoroughly investigated UE positioning performance in distributed MIMO systems. Based on the proposed EADF-based system model, we have derived FIM and PEB to characterize the maximum achievable one-shot positioning accuracy. The effects of UE tilt and spatial distributions of APs on the PEBs were evaluated. The numerical results highlight the performance gap between idealized and realistic EADF-based antenna models and indicate the importance of incorporating practical antenna properties into positioning analysis for accurate and realistic results. Furthermore, a comprehensive tracking framework has been developed by integrating a PHD filter with PEB-aware AP management. The effectiveness of the proposed framework has been validated using real-world distributed MIMO channel measurements. The results demonstrate that the framework achieves high tracking accuracy with low system overhead. This work offers valuable insights for practical AP deployments and further advancement of cooperative UE positioning techniques in distributed MIMO systems.

\section*{Acknowledgment}
The authors would like to thank Michiel Sandra for his assistance during the measurement campaign and for the valuable discussions.

\end{document}


%

\title{\textsc{\Huge SUPPLEMENTARY MATERIAL}
    {\large For}\\[0.75ex]
    {\Large Distributed MIMO Positioning: Fundamental Limit Analysis\\
    and User Tracking Framework Design}}

%

\author{Yingjie~Xu,
Xuesong~Cai, 
Ali~Al-Ameri,
        Sara~Willhammar, and~Fredrik~Tufvesson
}

%
%

\markboth{Journal of \LaTeX\ Class Files,~Vol.~xx, No.~x, December~2024}%
{Shell \MakeLowercase{\textit{et al.}}: Bare Demo of IEEEtran.cls for IEEE Journals}
%



\maketitle

%
\IEEEpeerreviewmaketitle
This supplementary material provides the derivations of the Fisher information~(FIM) matrix and the proofs of the theorems presented in the paper “\emph{Distributed MIMO Positioning: Fundamental Limit Analysis and User Tracking Framework Design}” by the same authors~\cite{Y. Xu2025}.
\section{Derivation of the Elements of $\mathbf{F}_{\Theta_k}$}
\label{AppendixA}
\vspace{-1mm}
The derivatives of $\mathbf{B}(\Theta_k)\cdot \gamma_k \triangleq \mathbf{D}$ with respect to the parameters $\Theta_k$ are given as 
\begin{equation}
\begin{split}
	\frac{\partial \mathbf{D} }{\partial \theta} = \sum_{\text{p}\in\{V,H\}} \alpha_{\text{pp}} \cdot c_{\rm{T}_{\text{p}}}' \left[\left(\mathbf{Q}_{\rm{R}_{\text{p}}} \cdot \left[\left(j\mathbf{\alpha}_{\theta} \odot \mathbf{w}_{\theta}\right) \otimes \mathbf{w}_{\phi}\right]\right) \otimes \mathbf{b}_f\right],
\end{split}
\label{eq: dD_dtheta}
\end{equation}
\begin{equation}
\begin{split}
	\frac{\partial \mathbf{D} }{\partial \phi} = \sum_{\text{p}\in\{V,H\}} \alpha_{\text{pp}} \cdot c_{\rm{T}_{\text{p}}}' \left[\left(\mathbf{Q}_{\rm{R}_{\text{p}}} \cdot \left[\mathbf{w}_{\theta} \otimes \left(j\mathbf{\alpha}_{\phi} \odot \mathbf{w}_{\phi}\right)\right]\right) \otimes \mathbf{b}_f\right],
\end{split}
\label{eq: dD_dphi}
\end{equation}
\begin{equation}
\begin{split}
	\frac{\partial \mathbf{D} }{\partial \tau} = \sum_{\text{p}\in\{V,H\}} \alpha_{\text{pp}} \cdot c_{\rm{T}_{\text{p}}}' \left[\left(\mathbf{Q}_{\rm{R}_{\text{p}}} \cdot \left[\mathbf{w}_{\theta} \otimes \mathbf{w}_{\phi}\right]\right) \otimes (-j2\pi\mathbf{f} \odot \mathbf{b}_f)\right],
\end{split}
\label{eq: dD_dtau}
\end{equation}
\begin{equation}
\frac{\partial \mathbf{D}}{\partial \mathcal{R}\{\alpha_{\rm{VV/HH}}\}} = c'_{\rm{T}_\text{V/H}} \left(\mathbf{b}_t \otimes [\mathbf{Q}_{\rm{R}_\text{V/H}} \cdot (\mathbf{w}_{\theta} \otimes \mathbf{w}_{\phi})] \otimes \mathbf{b}_f\right),
\label{eq: dD_dRalpha}
\end{equation}
\begin{equation}
\frac{\partial \mathbf{D}}{\partial \mathcal{I}\{\alpha_{\rm{VV/HH}}\}} = jc'_{\rm{T}_\text{V/H}} \left(\mathbf{b}_t \otimes [\mathbf{Q}_{\rm{R}_\text{V/H}} \cdot (\mathbf{w}_{\theta} \otimes \mathbf{w}_{\phi})] \otimes \mathbf{b}_f\right),
\end{equation}
where $\mathbf{f}=[f_1,...,f_{N_f}]^{\text{T}}$. The elements of $\mathbf{F}_{\Theta_k}$ can then be obtained by substituting the above equations into
$[\mathbf{F}_{\Theta_k}]_{i,j}
	= \frac{2}{\sigma^2} \mathcal{R}\left\{ \left(\frac{\partial \mathbf{B}(\mathbf{\Theta}_k) \cdot \gamma_k}{\partial \mathbf{\Theta}_{k,i}}\right)^{\text{H}} \cdot \frac{\partial \mathbf{B}(\mathbf{\Theta}_k) \cdot \gamma_k}{\partial \mathbf{\Theta}_{k,j}}\right\}.$

\section{Proof of Theorem 1}
\label{AppendixB}
For clarity, we omit the AP index $k$ in the following proof. Using the definition of equivalent FIM $\mathbf{F}_{\xi}=\mathbf{F}_{\xi\xi}-\mathbf{F}_{\xi\alpha}\mathbf{F}_{\alpha\alpha}\mathbf{F}_{\xi\alpha}^{\text{T}}$~\cite{Y. Shen2010_1}, it follows that $[\mathbf{F}_{\xi}]^{-1}=[\mathbf{F}_{\Theta}]^{-1}_{1:3,1:3}$ according to the Schur complement theory. Hence,
$\mathbf{F}_{\mathbf{p}}^{\text{Local}}=\left[[\mathbf{J}\mathbf{F}_{\xi}\mathbf{J}^{\text{T}}]^{-1}\right]^{-1}= \left[\mathbf{J}^{-\text{T}}[\mathbf{F}_{\Theta}]^{-1}_{1:3,1:3}\mathbf{J}^{-1}\right]^{-1},$
where 
\begin{equation}
\mathbf{J}^{-1}=\begin{bmatrix} 
\tau c\cos\theta\cos\phi  & -\tau c\cos\theta\sin\phi   & \sin\theta\cos\phi \\
\tau c\cos\theta\sin\phi   & \tau c\sin\theta\cos\phi  & \sin\theta\sin\phi \\  
-\tau c\sin\theta & 0 & \cos\theta
\end{bmatrix} ^{\text{T}}.
\end{equation}
Let $\mathbf{C}\triangleq\mathbf{J}^{-\text{T}}[\mathbf{F}_{\Theta}]^{-1}_{1:3,1:3}\mathbf{J}^{-1}$. Assuming that $[\mathbf{F}_\Theta]_{\Theta_i,{\Theta_j}}\ll[\mathbf{F}_\Theta]_{\Theta_i,{\Theta_i}}$, the entries of $\mathbf{C}$ are calculated as follows:
\begin{equation}
	C_{11}  \approx
	\frac{c^2\sin^2 \theta \cos^2 \phi}{[\mathbf{F}_{\Theta}]_{\tau\tau}} +
	\frac{c^2\tau^2 \sin^2 \phi \sin^2 \theta}{[\mathbf{F}_{\Theta}]_{\phi\phi}} +
	\frac{c^2\tau^2 \cos^2 \phi \cos^2 \theta}{[\mathbf{F}_{\Theta}]_{\theta\theta}},
\end{equation}
\begin{equation}
\begin{split}
		C_{12}=C_{21} \approx &\frac{\sin \phi \cos \phi}{[\mathbf{F}_{\Theta}]_{\theta\theta} [\mathbf{F}_{\Theta}]_{\phi\phi}[\mathbf{F}_{\Theta}]_{\tau\tau}}\big([\mathbf{F}_{\Theta}]_{\theta\theta} [\mathbf{F}_{\Theta}]_{\phi\phi}c^2 \sin^2 \theta \\
	&- [\mathbf{F}_{\Theta}]_{\theta\theta} [\mathbf{F}_{\Theta}]_{\tau\tau} c^2\tau^2 \sin^2 \phi + [\mathbf{F}_{\Theta}]_{\phi\phi} [\mathbf{F}_{\Theta}]_{\tau\tau} \\
    & \cdot c^2\tau^2 \cos^2 \phi),
\end{split}
\end{equation}
\begin{equation}
	C_{13}=C_{31} \approx
	\frac{([\mathbf{F}_{\Theta}]_{\theta\theta} -[\mathbf{F}_{\Theta}]_{\tau\tau} c^2\tau^2) \cos \phi \sin(2\theta)}{2 [\mathbf{F}_{\Theta}]_{\theta\theta}[\mathbf{F}_{\Theta}]_{\tau\tau}},
\end{equation}
\begin{equation}
	C_{22} \approx 
	\frac{c^2\sin^2 \theta \sin^2 \phi}{[\mathbf{F}_{\Theta}]_{\tau\tau}} +
	\frac{c^2\tau^2 \sin^2 \theta \cos^2 \phi}{[\mathbf{F}_{\Theta}]_{\phi\phi}} +
	\frac{c^2\tau^2 \sin^2 \phi \cos^2 \theta}{[\mathbf{F}_{\Theta}]_{\theta\theta}},
\end{equation}
\begin{equation}
	C_{23}=C_{32} \approx 
	\frac{-([\mathbf{F}_{\Theta}]_{\theta\theta} - [\mathbf{F}_{\Theta}]_{\tau\tau} c^2\tau^2) \sin \phi \sin(2\theta)}{2 [\mathbf{F}_{\Theta}]_{\theta\theta} [\mathbf{F}_{\Theta}]_{\tau\tau}},
\end{equation}
\begin{equation}
	C_{33} \approx 
	\frac{c^2\cos^2 \theta}{[\mathbf{F}_{\Theta}]_{\tau\tau}} + 
	\frac{c^2\tau^2 \sin^2 \theta}{[\mathbf{F}_{\Theta}]_{\theta\theta}}.
\end{equation}

It can be concluded that 
\begin{equation}
\begin{split}
\mathbf{C} =&\frac{(\tau c)^2}{[\mathbf{F}_{\Theta}]_{\theta\theta}} \mathbf{U}\left(\theta + \frac{\pi}{2}, \phi \right) + \frac{(\tau c \sin \theta)^2}{[\mathbf{F}_{\Theta}]_{\phi\phi}} \mathbf{U}\left(\frac{\pi}{2}, \phi + \frac{\pi}{2} \right)\\ 
& + \frac{c^2}{[\mathbf{F}_{\Theta}]_{\tau\tau}} \mathbf{U}(\theta, \phi),
\end{split}
\end{equation}
where $\mathbf{U}(\theta, \phi)$ is the 3D RDM. Then~(20) in~\cite{Y. Xu2025} can be easily obtained with $\mathbf{F}^{\text{Local}}=\mathbf{C}^{-1}$, which completes the proof.

\section{Proof of Theorem 3}
\label{AppendixD}
Rewrite the FIM in~(24) in~\cite{Y. Xu2025} as $\mathbf{F}_{\mathbf{P}}=\begin{bmatrix} F_{11}  & F_{12} \\ F_{21}  & F_{22} \end{bmatrix}$, where
\begin{equation}
F_{11} = \sum_{k=1}^{K} \frac{\left[ \mathbf{F}_{\Theta_k} \right]_{\phi\phi}}{c^2\tau_k^2} \sin^2 \phi_k' +\frac{\left[ \mathbf{F}_{\Theta_k} \right]_{\tau\tau}}{c^2} \cos^2 \phi_k',
\label{eq:F_11}
\end{equation}
\begin{equation}
F_{22} = \sum_{k=1}^{K} \frac{\left[ \mathbf{F}_{\Theta_k} \right]_{\phi\phi}}{c^2\tau_k^2} \cos^2 \phi_k' +\frac{\left[ \mathbf{F}_{\Theta_k} \right]_{\tau\tau}}{c^2} \sin^2 \phi_k',
\label{eq:F_22}
\end{equation}
\begin{equation}
F_{12}= F_{21}= \sum_{k=1}^{K}\left( \frac{\left[ \mathbf{F}_{\Theta_k} \right]_{\tau\tau}}{c^2}-\frac{\left[ \mathbf{F}_{\Theta_k} \right]_{\phi\phi}}{c^2\tau_k^2}\right) \cos \phi_k'\sin \phi_k',
\label{eq:F_12}
\end{equation}
where $\phi_k'=\phi_k+\omega_k$. Then 
\begin{equation}
\mathcal{P}({\mathbf{p}}^{\text{UE}})= \sqrt{\text{tr}\{\mathbf{F}_\mathbf{p}^{-1}\}}=\sqrt{\frac{F_{11}+F_{22}}{F_{11}F_{22}-(F_{12})^2}}.
\label{eq:PEB2}
\end{equation}
Inserting~(\ref{eq:F_11})--(\ref{eq:F_12}) into~(\ref{eq:PEB2}), we have
\begin{equation}
\begin{split}
&\mathcal{P}({\mathbf{p}}^{\text{UE}}) \\
&= \sqrt{\frac{2 \sum_{k}^{K} \left(F^{\tau\tau}_k + F^{\theta\theta}_k\right)}{\sum_{k, k'}^{K} \left( u_k u_{k'} \sin^2(\phi_k' - \phi_{k'}') +F^{\tau\tau}_{k'}F^{\theta\theta}_k+F^{\tau\tau}_kF^{\theta\theta}_{k'}\right)}}    
\end{split}
\label{eq:speb}
\end{equation}
where $F^{\tau\tau}_k=\frac{\left[ \mathbf{F}_{\Theta_k} \right]_{\tau\tau}}{c^2}$, $F^{\theta\theta}_k=\frac{\left[ \mathbf{F}_{\Theta_k} \right]_{\phi\phi}}{c^2\tau_k^2}$, and $u_k=\frac{\left[ \mathbf{F}_{\Theta_k} \right]_{\tau\tau}}{c^2}-\frac{\left[ \mathbf{F}_{\Theta_k} \right]_{\phi\phi}}{c^2\tau_k^2}$. We notice that 
\begin{equation}
\begin{split}
	\Lambda&\triangleq\sum_{k, k'}^{K} u_k u_{k'} \sin^2(\phi_k' - \phi_{k'}')\\
	&=    \frac{1}{2} \left[ \left| \sum_{k=1}^{K} u_k \right|^2 - \left| \sum_{k}^{K} u_k \exp(j2\phi_k') \right|^2 \right].  
\end{split}
\end{equation}
$\Lambda$ is maximized when $\left| \sum_{k}^{K} u_k \exp(j2\phi_k') \right|^2=0$. At this time, the PEB is minimized as 
\begin{equation}
\begin{split}
&\mathcal{P}({\mathbf{p}}^{\text{UE}}) \\
&= \sqrt{\frac{2 \sum_{k}^{K} \left(F^{\tau\tau}_k + F^{\theta\theta}_k\right)}{\frac{1}{2}\left| \sum_{k=1}^{K} u_k \right|^2+   \sum_{k, k'}^{K} \left( F^{\tau\tau}_{k'}F^{\theta\theta}_k+F^{\tau\tau}_kF^{\theta\theta}_{k'}\right)}}.    
\end{split}
\end{equation}
This can be easily rewritten to (26) in~\cite{Y. Xu2025}. Here we complete the proof.

\section{Proof of Theorem 4}
\label{AppendixE}
For clarity, we omit the user index $m$ in the following proof. We take the example of deriving $\mathbb{E}\left\{ \left[ \mathbf{F}_{\Theta_k} \right]_{\theta\theta}\right\}$. Applying the UE antenna model considered in Section~III-C, (12) in~\cite{Y. Xu2025} can be simplified as
\begin{equation}
\begin{split}
\left[ \mathbf{F}_{\Theta_k} \right]_{\theta\theta}
=& \frac{2E_s}{\sigma^2} \mathcal{R}\big\{\cos^2\beta(\mathbf{A}^{\theta}_{\rm{V}})^{\text{H}}\mathbf{A}^{\theta}_{\rm{V}}+\sin^2\beta(\mathbf{A}^{\theta}_{\rm{H}})^{\text{H}}\mathbf{A}^{\theta}_{\rm{H}}\\
&+\cos\beta\sin\beta\left((\mathbf{A}^{\theta}_{\rm{H}})^{\text{H}}\mathbf{A}^{\theta}_{\rm{V}}+(\mathbf{A}^{\theta}_{\rm{V}})^{\text{H}}\mathbf{A}^{\theta}_{\rm{H}}\right) \big\},
\label{eq:F_thetatheta3}
\end{split}
\end{equation}
where
$(\mathbf{A}^{\theta}_{\text{p}})^{\text{H}}\mathbf{A}^{\theta}_{\text{p}'}=\alpha_{\text{pp}}^{*}\alpha_{\text{p}'\text{p}'} \big( \left[(\mathbf{\alpha}_{\theta}\odot\mathbf{w}_{\theta})\otimes\mathbf{w}_{\phi}\right]\cdot \mathbf{Q}_{T_{\text{p}}}^{\text{H}}\mathbf{Q}_{T_{\text{p}'}}
\cdot\left[(\mathbf{\alpha}_{\theta}\odot\mathbf{w}_{\theta})\otimes\mathbf{w}_{\phi}\right]^{\text{H}}\big).$
When $\left(\mathbf{c}_{\rm{V}}\right) ^{\mathrm{H}}\mathbf{c}_{\rm{H}}+\left(\mathbf{c}_{\rm{H}}\right) ^{\mathrm{H}}\mathbf{c}_{\rm{V}}\ll \left(\mathbf{c}_{\rm{V}}\right) ^{\mathrm{H}}\mathbf{c}_{\rm{V}}+\left(\mathbf{c}_{\rm{H}}\right) ^{\mathrm{H}}\mathbf{c}_{\rm{H}}$, we have
$(\mathbf{A}^{\theta}_{\rm{H}})^{\text{H}}\mathbf{A}^{\theta}_{\rm{V}}+(\mathbf{A}^{\theta}_{\rm{V}})^{\text{H}}\mathbf{A}^{\theta}_{\rm{H}}\ll (\mathbf{A}^{\theta}_{\rm{H}})^{\text{H}}\mathbf{A}^{\theta}_{\rm{H}}+(\mathbf{A}^{\theta}_{\rm{V}})^{\text{H}}\mathbf{A}^{\theta}_{\rm{V}}.$
Therefore,
$\mathbb{E}\{ \left[ \mathbf{F}_{\Theta_k} \right]_{\theta\theta}\} \approx  \frac{2E_s}{\sigma^2} \mathcal{R}\bigg\{   \mathbb{E}\{ \cos^2\beta\} (\mathbf{A}^{\theta}_{\rm{V}})^{\text{H}}\mathbf{A}^{\theta}_{\rm{V}} 
+ \mathbb{E}\{ \sin^2\beta\} (\mathbf{A}^{\theta}_{\rm{H}})^{\text{H}}\mathbf{A}^{\theta}_{\rm{H}}\bigg\}.$
Since $\beta \sim \mathcal{VM}(\mu,\kappa)$, there is 
$\mathbb{E}[\cos(n\beta)] = \frac{I_n(k)}{I_0(k)} \cos(n\mu)$,
thus we have
\begin{equation}
\begin{aligned}
	\mathbb{E}[\cos^2 \beta] &= \frac{1}{2} \left( 1 + \mathbb{E}[\cos(2\beta)] \right) = \frac{1}{2} \left( 1 + \frac{I_2(\kappa)}{I_0(\kappa)} \cos(2\mu) \right), \\
	\mathbb{E}[\sin^2 \beta] &= \frac{1}{2} \left( 1 - \mathbb{E}[\cos(2\beta)] \right) = \frac{1}{2} \left( 1 - \frac{I_2(\kappa)}{I_0(\kappa)} \cos(2\mu) \right).
	\label{eq:E_cos^2beta}
\end{aligned}
\end{equation}
Then (29) in~\cite{Y. Xu2025} can be obtained with $z=z'=\theta$ by inserting~(\ref{eq:E_cos^2beta}) into~(\ref{eq:F_thetatheta3}). Other entries of $\mathbb{E}\left\{ \mathbf{F}_{\Theta_k}\right\}$ can be derived similarly, which completes the proof.
